\pdfoutput=1
\documentclass[11pt,a4paper]{article}
\usepackage{geometry}
\usepackage[pdftex]{graphicx}
\usepackage{color}
\usepackage{framed}
\usepackage[bottom]{footmisc}
\usepackage{amssymb}
\usepackage{latexsym}
\usepackage{amsmath}
\usepackage{mathtools}
\usepackage{enumitem}
\usepackage{fancyhdr}
\usepackage[hang,bf]{caption}
\usepackage[retainorgcmds]{IEEEtrantools}
\usepackage{bm}
\usepackage[bookmarks=true,pdfborder={0 0 0}]{hyperref}
\usepackage[all]{hypcap}
\newcommand{\bK}{\text{\bf{K}}}
\renewcommand{\Re}{\text{Re}}

\title{Hyperbolic monopoles from hyperbolic vortices}
\author{Rafael Maldonado\footnote{{\tt R.Maldonado@damtp.cam.ac.uk}}\vspace{0.2cm}\\\emph{Department of Applied Mathematics and Theoretical Physics,}\\\emph{Wilberforce Road, Cambridge CB3 0WA, U.K.}}

\begin{document}
\maketitle
\vspace{-7cm}
\begin{minipage}{\linewidth}
\begin{flushright}
DAMTP-2015-47
\end{flushright}
\end{minipage}
\vspace{6cm}
\begin{abstract}
\noindent Certain hyperbolic monopoles and all hyperbolic vortices can be constructed from $\text{SO}(2)$ and $\text{SO}(3)$ 
invariant Euclidean instantons, respectively.  This observation allows us to describe a large class of hyperbolic monopoles as 
hyperbolic vortices embedded into $\mathbb{H}^3$ and yields a remarkably simple relation between the two Higgs fields.  This 
correspondence between vortices and monopoles gives new insight into the geometry of the spectral curve and the moduli space of 
hyperbolic monopoles.  It also allows an explicit construction of the fields of a hyperbolic monopole invariant under a 
$\mathbb{Z}$ action, which we compare to periodic monopoles in Euclidean space.
\end{abstract}

\section{Introduction}
It has been known for some time that both the BPS monopole and the Abelian-Higgs vortex equations are more tractable in hyperbolic 
space (of a prescribed curvature) than in Euclidean space, with solutions expressible as rational functions.  The reason for this 
simplification is that both cases are reductions of the (conformally invariant) self-duality equations in $\mathbb{R}^4$, by an 
$\text{SO}(2)$ and an $\text{SO}(3)$ action respectively.

In this paper we explore the relation between monopoles and vortices in hyperbolic space.  In the remainder of this section we 
review the construction of hyperbolic monopoles and vortices.  Section \ref{symminst} discusses how the hyperbolic monopole and 
vortex equations come about from instanton reductions and shows how hyperbolic vortices can be used to construct hyperbolic 
monopoles.  A description of this procedure in terms of JNR data is given in section \ref{JNRsection}.  In section \ref{JNRSC} we 
look at the spectral curve of the resulting hyperbolic monopoles and compare to the spectral data of Euclidean monopoles.  The 
metric on the $2$-hyperbolic-monopole moduli space (defined via the connection at infinity) is compared to the physical metric on 
the underlying hyperbolic vortex moduli space in section \ref{secmod}.  Finally, in section \ref{chainssection} we use the 
procedure of section \ref{symminst} to construct a periodic hyperbolic monopole, for which a direct construction in terms of JNR 
or ADHM data is not currently known.

\subsection{Hyperbolic vortices}\label{hyperbolicvortices}
Abelian Higgs vortices consist of a complex Higgs field $\phi$ and a two-component gauge potential $a$.  At critical coupling 
there is a topological energy bound, and this fixes the number of zeros of $\phi$.  Away from its zeros, $|\phi|^2$ obeys the 
Taubes equation:
\begin{equation}
\Delta\log|\phi|^2+2(1-|\phi|^2)\,=\,0,\label{Taubeseq}
\end{equation}
where $\Delta$ is the Laplace-Beltrami operator, which for a conformally flat background is given by $\Omega^{-1}\nabla^2$, where 
$\Omega$ is the conformal factor and $\nabla^2$ is the Euclidean Laplacian.  On the hyperbolic plane of Gauss curvature 
$-1$ the Taubes equation can be reduced to the Liouville equation, which is integrable.  Working in the Poincar\'e disk model, 
solutions to the Taubes equation are given in terms of a holomorphic function $f(w)$ satisfying $|f(w)|\leq1$, with equality on 
the boundary of the disk $|w|=1$.  Explicitly,
\begin{equation}
\phi\,=\,\frac{1-|w|^2}{1-|f(w)|^2}\frac{df}{dw},\qquad\qquad a_{\bar{w}}\,=\,-\text{i}\partial_{\bar{w}}\log(\phi),\label{phifformula}
\end{equation}
which are defined up to a $\text{U}(1)$ gauge transformation.  For prescribed vortex locations it is possible in principle to 
construct the required function $f(w)$ as a Blaschke product.  An equivalent construction in terms of JNR data with poles on the 
boundary circle will be discussed in section \ref{JNRsection}.
\subsection{Hyperbolic monopoles}\label{intromonopoles}
$\text{SU}(2)$ monopoles consist of an adjoint-valued scalar $\Phi$ and a three-component gauge potential $A$.  In hyperbolic 
$3$-space $\mathbb{H}^3$ the Bogomolny monopole equations are
\begin{equation}
F_{ij}\,=\,\sqrt{\Omega}\,\epsilon_{ijk}D_k\Phi.\label{Bogomolnyeqs}
\end{equation}
Solutions are rational if the boundary condition $\|\Phi\|^2\coloneqq-\tfrac{1}{2}\text{tr}(\Phi^2)\to v^2$ has half-integer $v$.  
The simplest case has $v=\tfrac{1}{2}$, when a large family of monopoles can be constructed from JNR data with poles on the 
boundary of $\mathbb{H}^3$.  More generally, all solutions for $v=\tfrac{1}{2}$ arise from circle-invariant ADHM data, while for 
all half-integer $v$ one obtains a discrete version of the Nahm equations, known as the Braam-Austin equations \cite{BA90}.

Examples of $v=\tfrac{1}{2}$ hyperbolic monopoles which have been studied include those with axial \cite{Coc14} and Platonic 
\cite{MS14} symmetry.  More recently, monopoles of large charge have been modelled as magnetic bags \cite{BHS}.

\section{Symmetric instantons}\label{symminst}
The goal of this paper is to explore the relation between hyperbolic monopoles and vortices by means of the underlying 
$\text{SO}(3)$-invariant instanton.  We firstly lift the general vortex solution to an instanton using Witten's approach 
\cite{Wit77}, which is suited to the upper half space model of hyperbolic space.  The instanton is then reduced to a monopole by 
imposing a circle invariance.  This leads to a simple expression relating the monopole and vortex Higgs fields.  We then confirm 
that for this class of monopoles, the Bogomolny equations \eqref{Bogomolnyeqs} imply the Taubes equation \eqref{Taubeseq} on the 
vortex fields.
\subsection{Conformal rescalings}
Before we proceed, let us fix our conventions.  The metric on $\mathbb{E}^4$ is
\begin{equation*}
ds^2_{\mathbb{E}^4}\,=\,(dx^4)^2+(dx^1)^2+(dx^2)^2+(dx^3)^2\,=\,(dx^4)^2+(dx^1)^2+d\rho^2+\rho^2d\xi^2,
\end{equation*}
where $x^2=\rho\cos(\xi)$ and $x^3=\rho\sin(\xi)$.  Imposing independence from the coordinate $\xi$, this metric is 
conformally equivalent to hyperbolic $3$-space with the upper half space metric
\begin{equation}
ds^2_{\mathbb{H}^3}\,=\,\frac{1}{\rho^2}\left((dx^4)^2+(dx^1)^2+d\rho^2\right).\label{H3metric}
\end{equation}
Now introduce the coordinates $r\geq0$ and $\theta\in[0,\pi)$ via $x^1=r\cos(\theta)$, $\rho=r\sin(\theta)$.  Then $r$, $\theta$ and 
$\xi$ are standard spherical polar coordinates with respect to which
\begin{equation*}
ds^2_{\mathbb{E}^4}\,=\,(dx^4)^2+dr^2+r^2\left(d\theta^2+\sin^2(\theta)d\xi^2\right),
\end{equation*}
where $r^2=(x^1)^2+\rho^2=(x^1)^2+(x^2)^2+(x^3)^2$.  Quotienting by the angular dependence now gives a metric on the hyperbolic 
plane $\mathbb{H}^2$,
\begin{equation}
ds^2_{\mathbb{H}^2}\,=\,\frac{1}{r^2}\left((dx^4)^2+dr^2\right).\label{H2metric}
\end{equation}
The relation between the metrics \eqref{H2metric} and \eqref{H3metric} is interesting.  Restricting to $\theta=\pi/2$, 
\eqref{H2metric} reads
\begin{equation*}
\left.ds_{\mathbb{H}^2}^2\right|_{\theta=\frac{\pi}{2}}\,=\,\frac{1}{\rho^2}\left((dx^4)^2+d\rho^2\right),
\end{equation*}
which by comparison with \eqref{H3metric} is a slice of $\mathbb{H}^3$ (an equatorial slice of the unit ball model of 
$\mathbb{H}^3$ is a unit disc carrying a hyperbolic metric).  There is a more subtle reduction if we restrict to $\theta=0$.  Then 
\eqref{H2metric} becomes
\begin{equation}
\left.ds_{\mathbb{H}^2}^2\right|_{\theta=0}\,=\,\frac{1}{(x^1)^2}\left((dx^4)^2+(dx^1)^2\right).\label{boundarymetric}
\end{equation}
This is the boundary of $\mathbb{H}^3$ equipped with a hyperbolic metric, and is known as the `hemisphere model' of 
$\mathbb{H}^2$.  Since $x^1$ can take either sign, this is two copies of the hyperbolic plane glued along the $x^4$-axis.  By 
extension there is a family of such metrics, according to a choice of the angle $\theta$.

We will frequently use the ball model of $\mathbb{H}^3$, where cyclic symmetry is more apparent than in the upper half space 
model.  The ball model coordinates are given in terms of the upper half space coordinates by
\begin{equation*}
X^1+\text{i}X^2\,=\,\frac{2(x^4+\text{i}x^1)}{(x^1)^2+(x^4)^2+(\rho+1)^2},\qquad X^3\,=\,\frac{(x^1)^2+(x^4)^2+(\rho^2-1)}{(x^1)^2+(x^4)^2+(\rho+1)^2},
\end{equation*}
\begin{equation}
R^2\,=(X^1)^2+(X^2)^2+(X^3)^2\,=\,\frac{(x^1)^2+(x^4)^2+(\rho-1)^2}{(x^1)^2+(x^4)^2+(\rho+1)^2},\label{ballR}
\end{equation}
and the ball metric is
\begin{equation*}
ds^2\,=\,\frac{4}{(1-R^2)^2}\left((dX^1)^2+(dX^2)^2+(dX^3)^2\right). 
\end{equation*}
For completeness, we invert these expressions to give the upper half space coordinates in terms of the ball model coordinates:
\begin{equation*}
x^4+\text{i}x^1\,=\,\frac{2(X^1+\text{i}X^2)}{1+R^2-2X^3},\qquad\rho\,=\,\frac{1-R^2}{1+R^2-2X^3}.
\end{equation*}

\subsection{Dimensional reductions}\label{dimreds}
Monopoles and vortices in hyperbolic space are constructed by dimensional reductions of instantons on $\mathbb{E}^4$.  The 
self-duality (instanton) equations are conformally invariant, so solutions are unchanged under the conformal rescalings of the 
background metric described above.  Instantons invariant under a circle symmetry can then be dimensionally reduced to monopoles on 
$\mathbb{H}^3$, while $\text{SO}(3)$-invariant instantons give rise to hyperbolic vortices.
\par The reduction of circle-invariant instantons to hyperbolic monopoles was first considered by Atiyah \cite{Ati88} and carried 
out by Chakrabarti \cite{Cha86} and Nash \cite{Nas86}.  Given an instanton gauge potential $A_i(x^4,x^1,x^2,x^3)$ which is 
independent of $\xi=\tan^{-1}(x^3/x^2)$, one must perform a gauge transformation $G$ such that $A^G_i$ is explicitly 
independent of $\xi$.  In this gauge, the monopole Higgs field $\Phi$ is identified with $A^G_\xi$, and the monopole gauge 
potential has components $A^G_4$, $A^G_1$, $A^G_\rho$.
\par The relation between instantons and hyperbolic vortices first arose in Witten's search for cylindrically symmetric instantons 
\cite{Wit77}.  In the upper half plane model of $\mathbb{H}^2$, \eqref{H2metric}, a vortex consists of a Higgs field 
$\phi=\phi_1+\text{i}\,\phi_2$ and a gauge potential $a=a_4\,dx^4+a_r\,dr$, which we assume is in Coulomb gauge, 
$\partial_ia_i=0$.  From these one constructs an $\text{SO}(3)$-invariant instanton:
\begin{equation}
A_i\,=\,\frac{\text{i}}{2}\left(\frac{\phi_2+1}{r^2}\epsilon_{ijk}x^k\tau_j+\frac{\phi_1}{r^3}[r^2\tau_i-x^ix^j\tau_j]+\frac{a_rx^i}{r^2}\,x^j\tau_j\right),\qquad A_4\,=\,\frac{\text{i}a_4}{2r}\,x^j\tau_j,\label{instanton}
\end{equation}
%\begin{IEEEeqnarray}{rcl}
%_1&\,=&\,\frac{\text{i}}{2}\left(\frac{\phi_2+1}{r^2}[x^3\tau_2-x^2\tau_3]+\frac{\phi_1}{r^3}[((x^2)^2+(x^3)^2)\tau_1-x^1x^2\tau_2-x^1x^3\tau_3]+\frac{a_rx^1}{r^2}\,\bm{x}\cdot\bm{\tau}\right)\nonumber\\
%A_2&\,=&\,\frac{\text{i}}{2}\left(\frac{\phi_2+1}{r^2}[-x^3\tau_1+x^1\tau_3]+\frac{\phi_1}{r^3}[-x^1x^2\tau_1+((x^1)^2+(x^3)^2)\tau_2-x^2x^3\tau_3]+\frac{a_rx^2}{r^2}\,\bm{x}\cdot\bm{\tau}\right)\nonumber\\
%A_3&\,=&\,\frac{\text{i}}{2}\left(\frac{\phi_2+1}{r^2}[x^2\tau_1-x^1\tau_2]+\frac{\phi_1}{r^3}[-x^1x^3\tau_1-x^2x^3\tau_2+((x^1)^2+(x^2)^2)\tau_3]+\frac{a_rx^3}{r^2}\,\bm{x}\cdot\bm{\tau}\right)\nonumber\\
%A_4&\,=&\,\frac{\text{i}a_4}{2r}\,\bm{x}\cdot\bm{\tau},\label{instanton}
%\end{IEEEeqnarray}
where $i$ runs from $1$ to $3$, $r^2=(x^1)^2+(x^2)^2+(x^3)^2$ and all the $x^4$ dependence is encoded in the vortex fields.

We would like to explore the class of hyperbolic monopoles obtained by lifting hyperbolic vortices to instantons and then reducing 
by a circle action.  To do this, we first of all combine the $A_2$ and $A_3$ components of the instanton gauge potential 
\eqref{instanton} into $A_\rho=(x^2A_2+x^3A_3)/\rho$ and 
$A_\xi=-x^3A_2+x^2A_3$,
\begin{IEEEeqnarray*}{rcl}
A_\rho&\,=&\,\frac{\text{i}}{2}\left(\frac{\phi_2+1}{r^2}x^1[-\text{s}\tau_2+\text{c}\tau_3]+\frac{\phi_1x^1}{r^3}[-\rho\tau_1+x^1\text{c}\tau_2+x^1\text{s}\tau_3]+\frac{a_r\rho}{r^2}\,x^j\tau_j\right)\\
A_\xi&\,=&\,\frac{\text{i}}{2}\left(\frac{\phi_2+1}{r^2}\rho[\rho\tau_1-x^1\text{c}\tau_2-x^1\text{s}\tau_3]+\frac{\phi_1\rho}{r}[-\text{s}\tau_2+\text{c}\tau_3]\right),
\end{IEEEeqnarray*}
where $\text{c}=\cos(\xi)$ and $\text{s}=\sin(\xi)$.  Now the $A_i$ are rendered explicitly independent of $\xi$ by application of 
the gauge transformation
\begin{equation*}
A_i\,\mapsto\,A_i^G\,=\,G^{-1}A_iG+G^{-1}\partial_iG,
\end{equation*}
with $G=\text{exp}\left(-i\xi\tau_1/2\right)$.  The monopole fields are then simply the transformed gauge potential:
\begin{IEEEeqnarray}{rcl}
A_1^G&\,=&\,\frac{\text{i}}{2}\left(-\frac{\phi_2+1}{r^2}\rho\tau_3+\frac{\phi_1\rho}{r^3}[\rho\tau_1-x^1\tau_2]+\frac{a_rx^1}{r^2}[x^1\tau_1+\rho\tau_2]\right)\label{A1}\\
A_\rho^G&\,=&\,\frac{\text{i}}{2}\left(\frac{\phi_2+1}{r^2}x^1\tau_3-\frac{\phi_1x^1}{r^3}[\rho\tau_1-x^1\tau_2]+\frac{a_r\rho}{r^2}[x^1\tau_1+\rho\tau_2]\right)\label{Arho}\\
\Phi\,=\,A_\xi^G&\,=&\,\frac{\text{i}}{2}\left(\frac{\phi_2+1}{r^2}\rho[\rho\tau_1-x^1\tau_2]+\frac{\phi_1\rho}{r}\tau_3\right)-\frac{\text{i}}{2}\,\tau_1\label{Phi}\\
A_4^G&\,=&\,\frac{\text{i}a_4}{2r}[x^1\tau_1+\rho\tau_2].\label{A4}
\end{IEEEeqnarray}
The instanton fields \eqref{instanton} will match the standard JNR gauge introduced in section \ref{JNRsection} if
\begin{equation}
a_4\,=\,\frac{\phi_2+1}{r}\qquad\text{and}\qquad a_r\,=\,\frac{\phi_1}{r},\label{Mantongaugea}
\end{equation}
and with this choice the monopole gauge potential is automatically in Coulomb gauge, $d_iA_i^G=0$.

From \eqref{Phi} we obtain the key formula relating the norms of the vortex and monopole Higgs fields:
\begin{equation}
\|\Phi\|^2\,=\,\frac{\rho^2|\phi|^2+(x^1)^2}{4\,r^2},\label{modphisquared}
\end{equation}
where $r^2=(x^1)^2+\rho^2$ and $\phi$ is a function of $x^4$ and $r$.  Let us analyse this formula in more detail.  Recall that we 
are working on the upper half space whose boundary is the $(x^4,x^1)$ plane.  $\|\Phi\|^2$ has the correct boundary behaviour for 
a monopole with $v=\tfrac{1}{2}$ ($\|\Phi\|^2\to\tfrac{1}{4}$ as we approach the boundary $\rho\to0$), and its zeros occur 
where $x^1=0$ and $\phi=0$.  In the equatorial plane $x^1=0$, the monopole Higgs field $\|\Phi\|^2$ is proportional to the vortex 
Higgs field $|\phi|^2$, providing an obvious interpretation of the monopole as an embedded vortex.
\par
\begin{figure}
\centering
\includegraphics[width=0.8\linewidth]{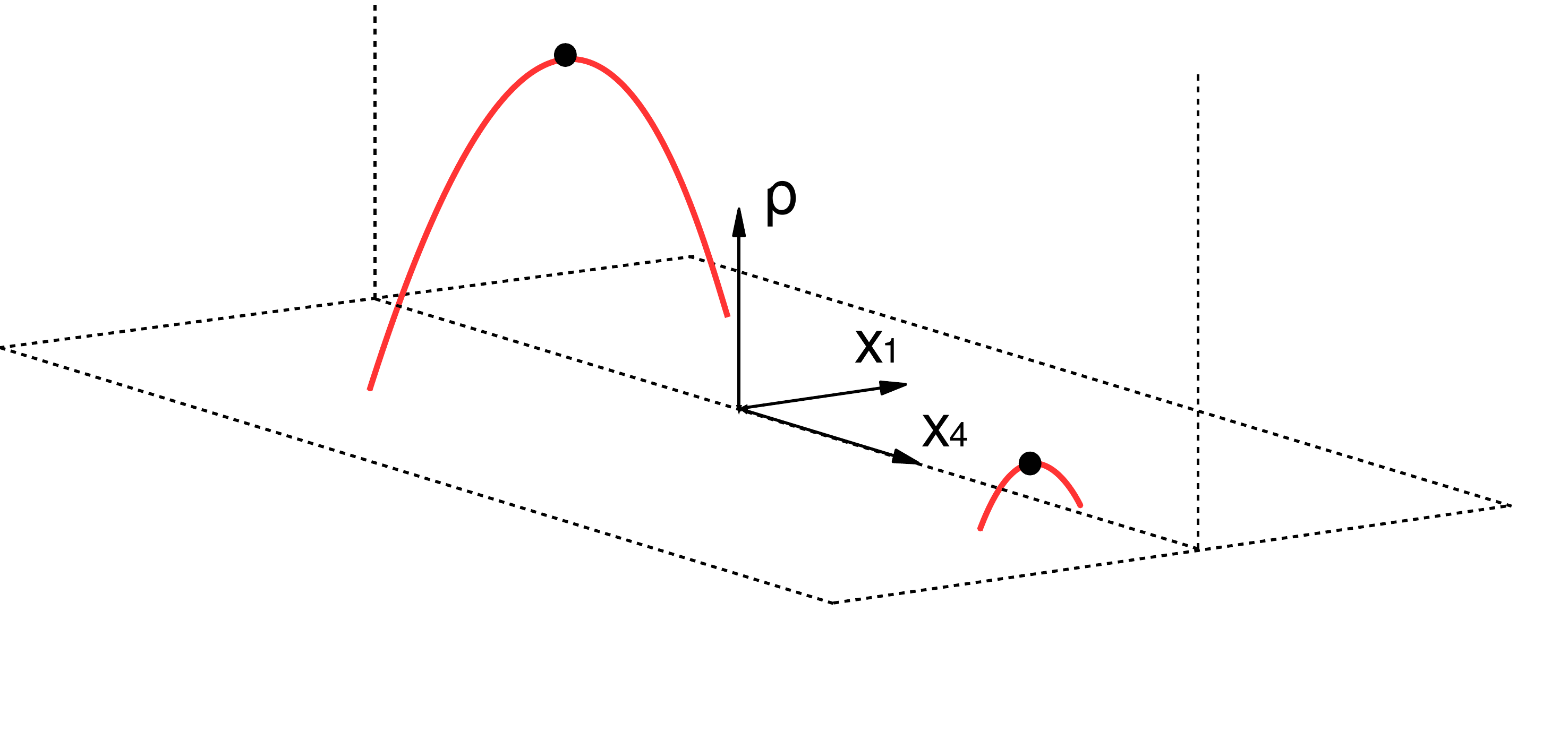}\vspace{-1cm}
\caption{Monopoles are located on the black dots in the plane $x^1=0$.  The vortex Higgs field $\phi$ vanishes along the solid 
lines, representing geodesics in the upper half space.}\label{fig1}
\end{figure}

 Now take $(x^4_0,r_0)$ to be the position of a vortex zero.  Setting $r=r_0$ defines a geodesic in the upper half space:~a 
semicircle which meets the boundary at $(x^4,x^1)=(x^4_0,\pm r_0)$, as shown in figure \ref{fig1}.  As a function of the 
hyperbolic distance $d_{\text{H}}$ from the monopole zero, measured along this geodesic, the Higgs field is
\begin{equation}
\|\Phi\|^2_{\phi=0}\,=\,\frac{(x^1)^2}{4r^2}\,=\,\frac{1}{4}\tanh^2(d_{\text{H}}).\label{Phiphi0}
\end{equation}
This is precisely the radial profile function of a single hyperbolic monopole, and the result \eqref{Phiphi0} is independent of 
the multiplicity of the associated monopole zero and of its position relative to any other monopoles in the 
configuration.\footnote{The energy density, which depends on derivatives of $\|\Phi\|^2$, is proportional to the radial energy 
density profile of a single hyperbolic monopole.  The constant of proportionality depends on the leading behaviour of $\|\Phi\|^2$ 
near its zeros.}  In section \ref{JNRSC} we will see that these distinguished geodesics are always spectral lines.

A similar analysis to that given in this section allowed Cockburn \cite{Coc14} to relate axially symmetric monopoles to charge one 
monopoles of half-integer mass $v>\tfrac{1}{2}$.

\subsection{Field equations}\label{fieldequations}
Now let us check that the Bogomolny equations imply the vortex equations.  Note that 
$\partial_1\tilde{\phi}=x^1(\partial_r\tilde{\phi})/r$, $\partial_\rho\tilde{\phi}=\rho(\partial_r\tilde{\phi})/r$, where 
$\tilde{\phi}$ represents any of the vortex fields, which are independent of $\theta$.  Then, using the fields 
\eqref{A1}-\eqref{A4} but suppressing the superscript ${}^G$ for clarity,
%\begin{IEEEeqnarray*}{rcl}
%F_{41}^G&\,=&\,\frac{\text{i}}{2}\left(-\frac{\rho}{r^2}\partial_4\phi_2\tau_3+\frac{\rho}{r^3}\partial_4\phi_1[\rho\tau_1-x^1\tau_2]+\frac{x^1}{r^2}\partial_4a_r[x^1\tau_1+\rho\tau_2]\right)\nonumber\\
%&&-\frac{\text{i}}{2}\left(\frac{x^1}{r^2}\partial_ra_4[x^1\tau_1+\rho\tau_2]+\frac{a_4\rho}{r^3}[\rho\tau_1-x^1\tau_2]\right)\nonumber\\
%&&+\frac{\text{i}a_4}{2r}\left(\frac{\phi_2+1}{r^2}\rho[\rho\tau_1-x^1\tau_2]+\frac{\phi_1\rho}{r}\tau_3\right),\nonumber
%\end{IEEEeqnarray*}
%\begin{IEEEeqnarray*}{rcl}
%D_\rho^G\Phi&\,=&\,\frac{\text{i}}{2}\left(\frac{\rho^2}{r^3}\partial_r\phi_2[\rho\tau_1-x^1\tau_2]+2\frac{\phi_2+1}{r^4}\rho x^1[x^1\tau_1+\rho\tau_2]+\frac{1}{r^3}[\rho^2r\partial_r\phi_1+(x^1)^2\phi_1]\tau_3\right)\\
%&&-\frac{\text{i}}{2}\left(\frac{(\phi_2+1)^2}{r^4}\rho x^1[x^1\tau_1+\rho\tau_2]-\frac{\phi_2+1}{r^2}x^1\tau_2+\frac{\phi_1^2\rho x^1}{r^4}[x^1\tau_1+\rho\tau_2]+\frac{\phi_1(x^1)^2}{r^3}\tau_3\right)\\
%&&-\frac{\text{i}}{2}\left(-\frac{\phi_2+1}{r^2}a_r\rho^2\tau_3+\frac{\phi_1a_r\rho^2}{r^3}[\rho\tau_1-x^1\tau_2]+\frac{a_r\rho^2}{r^2}\tau_3\right),
%\end{IEEEeqnarray*}
%which satisfy the hyperbolic version of the Bogomolny equations,
%\begin{equation*}
%F_{ij}^G\,=\,\frac{1}{\rho}\,\epsilon_{ijk}D_k^G\Phi,
%\end{equation*}
%with $\epsilon_{41\rho}=1$.  Simplifying and gathering terms,
\begin{IEEEeqnarray}{rcl}
F_{41}\,&=&\,\partial_4A_1-\partial_1A_4+[A_4,A_1]\nonumber\\&=&\,\frac{\text{i}}{2r^2}\left(\frac{\rho}{r}\left(\partial_4\phi_1+a_4\phi_2\right)\tau_a+x^1\left(\partial_4a_r-\partial_ra_4\right)\tau_b-\rho\left(\partial_4\phi_2-a_4\phi_1\right)\tau_3\right)\label{F41}
\end{IEEEeqnarray}
\begin{IEEEeqnarray*}{rcl}
D_\rho\Phi\,&=&\,\partial_\rho\Phi+[A_\rho,\Phi]\nonumber\\&=&\,\frac{\text{i}\rho}{2r^2}\left(\frac{\rho}{r}\left(\partial_r\phi_2-a_r\phi_1\right)\tau_a+\frac{x^1}{r^2}\left(1-\phi_1^2-\phi_2^2\right)\tau_b+\rho\left(\partial_r\phi_1+a_r\phi_2\right)\tau_3\right),
\end{IEEEeqnarray*}
where $\tau_a=[\rho\tau_1-x^1\tau_2]$ and $\tau_b=[x^1\tau_1+\rho\tau_2]$.  The Bogomolny equations
\begin{equation*}
F_{ij}\,=\,\frac{1}{\rho}\,\epsilon_{ijk}D_k\Phi
\end{equation*}
with $\epsilon_{41\rho}=1$ imply the complex vortex equation $(D^{(\text{v})}_4+\text{i}D^{(\text{v})}_r)\phi=0$, where the 
superscript ${}^{(\text{v})}$ is used to distinguish the covariant deriviative for the vortex fields, 
$D_i^{(\text{v})}=\partial_i-\text{i}a_i$, from its counterpart for monopole fields.  One similarly obtains the real vortex 
equation, $r^2B=(1-\phi_1^2-\phi_2^2)$.
%Note that the Bogomolny equation $\rho F_{1\rho}^G=D_4^G\Phi$ only implies the complex vortex equation, and not the real one.
It would be interesting to study whether a similar embedding of the vortex equations into $\mathfrak{su}(2)$ is be possible for 
$v\neq\tfrac{1}{2}$.

\subsection{Monopole number}
We should check that the fields \eqref{A1}-\eqref{A4} have the correct topology to be monopoles.  The fields are 
reflection-symmetric:~the replacement $x^1\to-x^1$ reverses the orientation and changes the fields by a gauge:
\begin{equation*}
A_1'\,=\,-\tau_2A_1\tau_2,\qquad A_\rho'\,=\,\tau_2A_\rho\tau_2,\qquad\Phi'\,=\,-\tau_2\Phi\tau_2,\qquad A_4'\,=\,\tau_2A_4\tau_2.\label{orientation}
\end{equation*}
Now compute the Chern number by performing the integral
\begin{equation}
c_1\,=\,-\int\frac{\text{tr}(F\Phi)}{4\pi\|\Phi\|}\label{chern}
\end{equation}
over the boundary of $\mathbb{H}^3$.  In upper half space coordinates this is the $x^4-x^1$ plane.  Setting $\rho=0$, we have from \eqref{Phi} 
and \eqref{F41} that
\begin{equation*}
\Phi^0\,=\,-\frac{\text{i}}{2}\,\tau_1,\qquad\qquad F_{41}^0\,=\,\frac{\text{i}}{2}\left(\partial_4a_r-\partial_ra_4\right)\tau_1\,=\,\frac{\text{i}}{2}\,B\,\tau_1.
\end{equation*}
Evaluating \eqref{chern}, we get
\begin{equation}
c_1\,=\,-\frac{1}{2\pi}\int_{\partial\mathbb{H}^3}\frac{1}{2}\,B\,dx^4\wedge dx^1\label{integral}
\end{equation}
Now $r^2=(x^1)^2$ (since $\rho=0$), but unlike the coordinate $r$, $x^1$ spans the entire real line and the integral 
\eqref{integral} is performed over two copies of the upper half plane.  This gives
\begin{equation*}
c_1\,=\,-\frac{1}{2\pi}\int_{\mathbb{H}^2}\frac{1}{2}\,\frac{B}{\Omega}\,\left(2\,\Omega\,dx^4\wedge dr\right)\,=\,-N,
\end{equation*}
so a charge $N$ vortex lifts to a charge $N$ monopole.
%This also works if we use $\ast D\Phi$ in place of $F$:~the divergent term proportional to $\tau_2$ is removed when projecting along $\Phi$.  One might also have to worry about any contributions from the point at infinity ($\rho\to\infty$).  The gauge field $F_{41}^G$ vanishes in this limit, so the hope is that it does not contribute.

\subsection{Energy density}\label{energysection}
The energy density of a monopole is obtained by applying the Laplace-Beltrami operator to $\|\Phi\|^2$.  In the upper-half-space 
model we are using, this is
\begin{equation*}
\mathcal{E}\,=\,\rho^2\left(\partial_1^2+\partial_4^2+\partial_\rho^2-\frac{1}{\rho}\,\partial_\rho\right)\|\Phi\|^2.
\end{equation*}
Written in terms of derivatives of the vortex Higgs field $|\phi|^2$ gives
\begin{equation}
\mathcal{E}\,=\,\frac{\rho^4}{r^4}\,\left(\frac{1}{4}\,\Delta|\phi|^2+\frac{1}{2}\,(1-|\phi|^2)\right),\qquad\text{where}\qquad\Delta\,=\,r^2\left(\partial_4^2+\partial_r^2\right)\label{energy}
\end{equation}
is the Laplace-Beltrami operator acting on the vortex Higgs field in the upper half plane model of $\mathbb{H}^2$.  We recognise 
the bracketed term in \eqref{energy} as the energy density of the vortex defined in \cite{MN99}.  Integrating over the upper half 
space we find
\begin{equation*}
E\,=\,\int\mathcal{E}\,\frac{1}{\rho^3}\,d\rho\,dx^1\,dx^4\,=\,2\pi N.
%&=&\,\frac{1}{4}\int\sin(\theta)\left(\partial_4^2+\partial_r^2\right)\left(\text{e}^h-h\right)dr\,d\theta\,dx^4\\
%&=&\,2\pi N.\phantom{\int}
\end{equation*}
%The factors of $\pi$ are a bit of a headache, but the important thing is that the total energy is proportional to $N$.  Alternatively, completing the square demonstrates positivity of the energy (and shows that this is really the Bogomolny energy of YMH vortices):
%The total energy is bounded by a Bogomolny argument:
%\begin{equation}
%\mathcal{E}\,=\,\frac{\rho^4}{4r^4}\left(r^2\text{e}^h\left((\partial_4h)^2+(\partial_rh)^2\right)+2\left(1-\text{e}^h\right)^2\right)\,\geq\,0.\label{energybound}
%\end{equation}

%\\ \\
%Consider a vortex at the origin of the Poincar\'e disk with the circularly symmetric expansion
%\begin{equation*}
%|\phi|^2\,=\,c^2+K(\delta|w|)^2+\dots\,=\,c^2+\tfrac{1}{4}K\left((\delta x^4)^2+(\delta r)^2\right)+\dots.
%\end{equation*}
%The resulting monopole energy density at this point is
%\begin{equation}
%\mathcal{E}\,=\,\tfrac{1}{2}(1-c^2)+\tfrac{1}{4}K.\label{Eatmin}
%\end{equation}
%For $c\neq0$ (i.e.~if $|\phi|^2$ attains a local maximum) it is more convenient to use \eqref{energybound} to write
%\begin{equation}
%\mathcal{E}\,=\,\tfrac{1}{2}\left(1-c^2\right)^2,\qquad\qquad(c\,\neq\,0)\label{Eatmax}
%\end{equation}
%as this does not require computation of $K$.  Note that this does not work when $c=0$, as then $h\to-\infty$ at the vortex 
%osition and there is a contribution from the first term in \eqref{energybound}.

\subsection{Example:~a single monopole}
Let us illustrate our discussion with a simple example.  A single vortex in the Poincar\'e disk has
\begin{equation}
|\phi|\,=\,\frac{2|w|}{1+|w|^2}.\label{onevortex}
\end{equation}
Now convert to upper half plane coordinates $z=x^4+\text{i}r$ using
\begin{equation*}
w\,=\,\frac{\text{i}-z}{\text{i}z-1},
\end{equation*}
This gives
%\begin{equation*}
%\phi_1\,=\,\frac{2x^4}{(x^4)^2+r^2+1}\qquad\text{and}\qquad\phi_2\,=\,\frac{(x^4)^2+r^2-1}{(x^4)^2+r^2+1},
%\end{equation*}
%so
\begin{equation}
|\phi|^2\,=\,\frac{((x^4)^2+(r-1)^2)((x^4)^2+(r+1)^2)}{((x^4)^2+r^2+1)^2},\label{oneuhpvortex}
\end{equation}
then from \eqref{modphisquared} and \eqref{ballR} we find, after some manipulation,
\begin{equation}
\|\Phi\|^2\,=\,\frac{1}{4}-\frac{\rho^2}{\left((x^1)^2+(x^4)^2+\rho^2+1\right)^2}\,=\,\frac{R^2}{(1+R^2)^2}.\label{onemonopole}
\end{equation}
Applying the energy density formula \eqref{energy} to the vortex \eqref{onevortex} gives
\begin{equation*}
\mathcal{E}_1\,=\,\frac{3}{2}\left(\frac{1-R^2}{1+R^2}\right)^4
\end{equation*}
as expected for the charge one monopole \eqref{onemonopole}.  In section \ref{chainssection} we will use this method to obtain a 
new explicit hyperbolic monopole solution.

%More generally, the formula \eqref{modphisquared} allows us to construct an axial $N$-monopole from the circularly symmetric vortex
%\begin{equation*}
%\phi\,=\,\frac{(N+1)w^N(1-|w|^2)}{1-|w|^{2(N+1)}}.
%\end{equation*}
%The result matches the axial hyperbolic monopole obtained by Cockburn \cite{Coc14} using a different approach.

\section{JNR construction}\label{JNRsection}
The JNR Ansatz \cite{JNR77} gives a large class of instantons.  An $N$-instanton is generated by the function (harmonic in 
$\mathbb{R}^4$)
\begin{equation*}
\psi\,=\,\sum_{j=0}^N\frac{\lambda_j^2}{|x-\gamma_j|^2}
\end{equation*}
which gives the instanton gauge potentials
\begin{equation*}
A_i\,=\,\frac{\text{i}}{2}\left[\epsilon_{ijk}\,\partial_j\log(\psi)\tau_k+\partial_4\log(\psi)\tau_i\right],\qquad A_4\,=\,-\frac{\text{i}}{2}\,\partial_i\log(\psi)\tau_i
\end{equation*}
where $\tau_i$ are the Pauli matrices.  Only for $N=1$ does the JNR construction give all possible instantons.

The dimensional reductions of the preceding section can be made at the level of JNR data.  Circle-invariant JNR data gives a 
subset of hyperbolic monopoles.  The poles of $\psi$ must lie on a plane (the fixed set of a circle action) in $\mathbb{E}^4$, 
which becomes the boundary of $\mathbb{H}^3$.  Counting parameters suggests that all hyperbolic monopoles for $N\leq3$ can be 
generated in this way \cite{BCS15}.  To reduce the monopoles to vortices, we have the additional constraint that the poles must be 
on the fixed set of an $\text{SO}(3)$ action, i.e.~on a line in $\mathbb{E}^4$.  It was shown by Manton \cite{Man78} that the JNR 
Ansatz generates all hyperbolic vortices, i.e.~that it is gauge-equivalent to the formulation of section \ref{hyperbolicvortices}.

A suitable definition of `centered' hyperbolic monopoles is given in \cite{MNS03}.  The centered moduli space has dimension 
$4(N-1)$, while the moduli space of centered hyperbolic vortices has dimension $2(N-1)$.  There is an $S^2$ worth of freedom in 
our choice of embedding of the hyperbolic vortices into $\mathbb{H}^3$, so the construction presented in this paper gives a $2N$ 
dimensional family of centered hyperbolic monopoles.  In particular, we obtain all centered $2$-monopoles, whose moduli space is 
explored in section \ref{secmod}.

Using the same upper half space coordinates as before, a monopole Higgs field is constructed using the JNR function
\begin{equation}
\psi\,=\,\sum_{j=0}^N\frac{\lambda_j^2}{|x^4+\text{i}x^1-\gamma_j|^2+\rho^2}\label{JNRpsi}
\end{equation}
in
\begin{equation}
\|\Phi\|^2\,=\,\frac{\rho^2}{4\psi^2}\left(\left(\frac{\partial\psi}{\partial x^4}\right)^2+\left(\frac{\partial\psi}{\partial x^1}\right)^2+\left(\frac{\psi}{\rho}+\frac{\partial\psi}{\partial\rho}\right)^2\right).\label{JNRPhi}
\end{equation}
Placing all the poles of \eqref{JNRpsi} on the real $x^4$-axis gives
\begin{equation}
\psi\,=\,\sum_{j=0}^N\frac{\lambda_j^2}{(x^4-\gamma_j)^2+r^2},\label{JNRpsi2}
\end{equation}
and the vortex Higgs field is given by \cite{Man78}
\begin{equation}
|\phi|^2\,=\,\frac{r^2}{\psi^2}\left(\left(\frac{\partial\psi}{\partial x^4}\right)^2+\left(\frac{\psi}{r}+\frac{\partial\psi}{\partial r}\right)^2\right)\,=\,-r^2\left(\partial_4^2+\partial_r^2\right)\log(r\psi).\label{vortexJNRformula}
\end{equation}
Fixing the phase of $\phi$ by specialising to Coulomb gauge and using the relations \eqref{Mantongaugea} gives the components of 
the gauge potential as
\begin{equation}
a_4\,=\,-\partial_r\log\psi,\qquad\qquad a_r\,=\,\partial_4\log\psi.\label{Mantongaugeb}
\end{equation}

Using \eqref{JNRpsi2} in \eqref{JNRPhi} and changing variables again gives the relation \eqref{modphisquared}.  Of course, there 
are certain vortex configurations for which the JNR function $\psi$ is not known.  The more general argument of section 
\ref{symminst} ensures that \eqref{modphisquared} is still valid, and it is for these configurations that the construction of 
monopoles as an embedding of vortices provides truly novel monopole solutions.

The remarkable similarity between \eqref{JNRPhi} and \eqref{vortexJNRformula} invites us to consider a further dimensional 
reduction.  The resulting one-dimensional field theory describes the $\mathrm{SO}(4)$-invariant instanton.  Using the radial 
coordinate $\varrho^2=r^2+(x^4)^2$ we define\footnote{There is, of course, only one $\text{SO}(4)$ symmetric 't Hooft function, 
namely $\psi=1+\lambda^2/\varrho^2$, but we will stick to using $\psi$ in order to highlight the analogy with the previous 
reduction from $3$ to $2$ dimensions.}
\begin{equation}
\varphi^2\,=\,\frac{\varrho^2}{\psi^2}\left(\frac{\psi}{\varrho}+\frac{d\psi}{d\varrho}\right)^2,\label{kinkphi}
\end{equation}
where $\psi$ is a function of $\varrho$ only.  Combining \eqref{kinkphi} with \eqref{vortexJNRformula}, the corresponding vortex Higgs field is
\begin{equation}
|\phi|^2\,=\,\frac{r^2\varphi^2+(x^4)^2}{r^2+(x^4)^2}.\label{vortexkink}
\end{equation}
Mimicking what we did in section \ref{fieldequations}, we substitute \eqref{vortexkink} into the Taubes equation \eqref{Taubeseq}, 
to yield
\begin{equation*}
\frac{d\varphi}{d\,\log(\varrho)}\,=\,1-\varphi^2,
\end{equation*}
which is the Bogomolny equation for a $\varphi^4$ kink.  In other words, we can obtain the charge $1$ hyperbolic vortex 
\eqref{oneuhpvortex} by embedding the (essentially unique) $\varphi^4$ kink into $\mathbb{H}^2$.  Lifting to $\mathbb{H}^3$, the 
hyperbolic tangent function describing the $\varphi^4$ kink shows up when the Higgs field of a single monopole is expressed as a 
function of hyperbolic distance from the Higgs zero, \eqref{Phiphi0}.  By a change of coordinates we regain the BPST instanton 
\cite{BPST75}.  The $\varrho$ coordinate of the kink is precisely the scale size $\lambda$ of the instanton.

%For a single monopole this result is true for any geodesic through the monopole.  All $N=2$ monopoles 
%can be constructed from JNR poles on an equator of $\mathbb{H}^3$, and the direction of these geodesics picks out an orientation.  
%For $N\geq3$ these lines only exist if the monopole configuration is an embedded vortex.
\section{Spectral data}\label{JNRSC}
The spectral curve of a hyperbolic monopole is defined by scattering data, as the set of geodesics along which
\begin{equation}
(D_s-\text{i}\Phi)w\,=\,0\label{specdef}
\end{equation}
has normalisable solutions, where $s$ is the arc length along the curve.  The spectral curve can be given explicitly in terms of 
the positions and weights of JNR poles \cite{BCS15}:
\begin{equation}
\mathcal{S}:\qquad\sum_{j=0}^N\lambda_j^2\prod_{k\neq j}(\zeta-\gamma_k)(1+\eta\bar{\gamma}_k)\,=\,0.\label{scjnr}
\end{equation}
Geodesics in $\mathbb{H}^3$ are parametrised in terms of their endpoints $\zeta$ and $-\bar{\eta}^{-1}$ on the boundary 
$\mathbb{R}^2\cong\mathbb{C}$.  We are interested in embedded vortices, where all JNR poles lie on the real axis, so 
$\gamma_k=\bar{\gamma}_k$.  Any $2$-monopole can be cast in this form by an appropriate choice of centre and orientation.

In the following sections we study three distinguished classes of spectral lines.

\subsection{Spectral lines through the monopole zeros}\label{speclinesthroughzeros}
Consider a vortex configuration embedded in $\mathbb{H}^3$ as described in section \ref{dimreds}, where it was observed that 
geodesics through monopole zeros orthogonal to the plane $x^1=0$ have $\phi=0$.  The monopole Higgs field $\Phi$ along this line 
is the radial field of a unit charge hyperbolic monopole.  It then follows from the definition \eqref{specdef} that such geodesics 
are spectral lines, by virtue of the fact that all spectral lines of a charge $1$ monopole pass through the zero.  We see this by 
expressing $\phi$ in terms of JNR data, such that
\begin{equation*}
\phi(z_0)\,=\,0\qquad\Rightarrow\qquad\left.\left((\bar{z}-z)\partial_z\log(\psi)\right)\right|_{z=z_0}\,=\,1\qquad\Rightarrow\qquad\sum_{j=0}^N\lambda_j^2\prod_{k\neq j}(z_0-\gamma_k)^2\,=\,0,
\end{equation*}
where $z=x^4+\text{i}r$.  Solutions for $z=z_0$ define geodesics in $\mathbb{H}^3$ which meet the boundary of the upper half space 
at $\zeta=z_0$ and $\zeta=\bar{z}_0$.  By comparison with \eqref{scjnr}, we see that these geodesics are in fact the unique 
spectral lines with $\eta=-\zeta^{-1}$, i.e.~which intersect the plane $x^1=0$ at right angles.  This observation should be 
contrasted with the case of Euclidean monopoles, when the spectral lines of a generic charge $2$ monopole only approximately 
pinpoint the zero.
%Note that in general there can be more spectral lines through the monopole zeros.  From \eqref{scjnr} it is clear that geodesics 
%between any pair of JNR poles are also spectral lines, and thus the cyclic monopole with odd $N$ has an infinite number of 
%spectral lines through its centre.  Nevertheless, for $N>1$ the distinguished spectral line described above is believed to be the only 
%spectral line through the monopole centre which is not contained in the plane $x^1=0$.

\subsection{Spectral lines in the plane of the vortices}
We now analyse some of the spectral lines described by \eqref{scjnr}.  Firstly, note that geodesics between any pair of JNR poles 
are spectral lines.  It is also clear that there are precisely $N$ spectral lines for each choice of $\zeta$ on the boundary, and 
that any geodesic with $\zeta\in\mathbb{R}$ also has $\eta\in\mathbb{R}$.  Specialising to $N=2$ with $\zeta\in\mathbb{R}$ leads 
to an interesting geometric picture in terms of Poncelet's theorem, which has already given insight into the geometry of 
instantons \cite{Har78} and indeed hyperbolic monopoles \cite{Hit95}.  We will work through the details explicitly in our case, 
making use of various theorems of Daepp-Gorkin-Mortini \cite{DGM02} and D.~Singer \cite{Sin06}.

We work with the ball model of $\mathbb{H}^3$, where the equatorial slice defined by $\zeta\in\mathbb{R}$ is a Poincar\'e disk 
with complex coordinate $w=X^1+\text{i}X^3$.  The boundary $w=\text{i}\,\text{e}^{-\text{i}\theta}$ is related to the coordinate 
$\zeta$ by stereographic projection:~$\zeta=\cot(\theta/2)$.  For notational convenience we will consider a centered $2$-monopole 
aligned with the $X^3$-axis, although the discussion follows through for any value of the (vortex) moduli.  The spectral curve can 
be parametrised as
\begin{equation}
\gamma^2(\zeta^2-\gamma^2)(1-\eta^2\gamma^2)-(1-\gamma^4)\eta(\zeta-\eta\gamma^2)\,=\,0,\label{speccharge2}
\end{equation}
with $\tfrac{1}{3}\!\leq\!\gamma^2<1$, and the relation between $\gamma$ and the monopole separation will be clarified in section 
\ref{secmod}.

\begin{figure}
\centering
\includegraphics[width=0.4\linewidth]{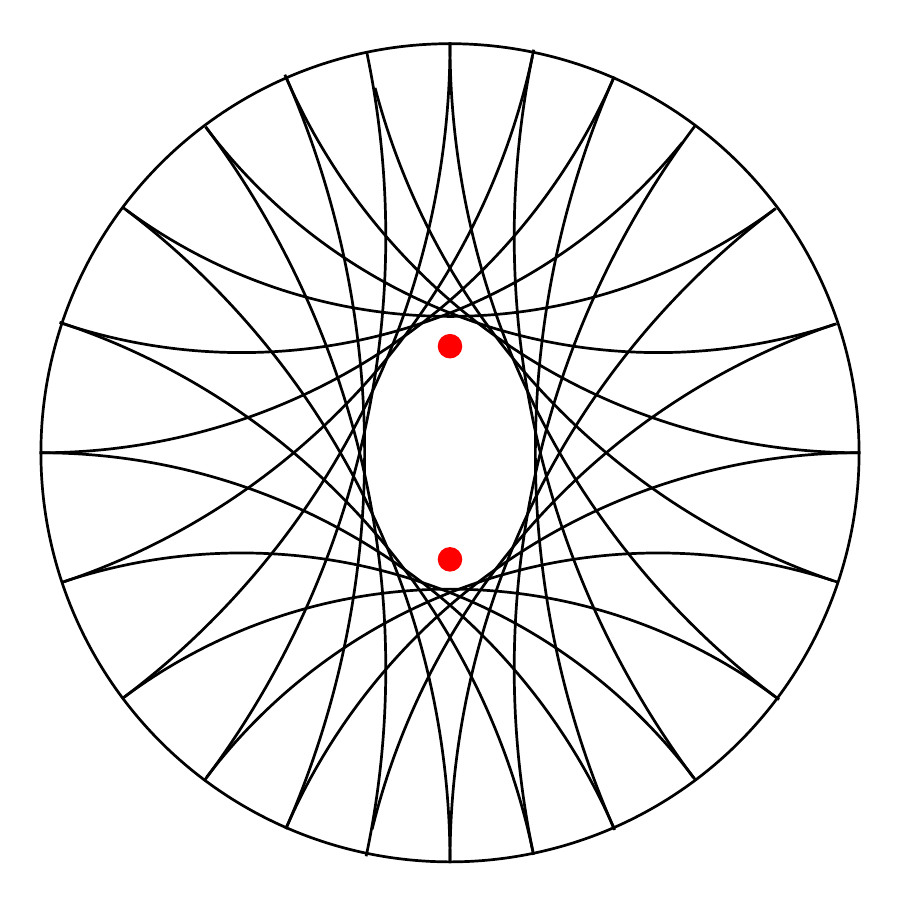}
\caption{Some spectral lines for a charge $2$ hyperbolic monopole with $\gamma^2=\tfrac{1}{4}$, restricted to the equatorial plane 
$X^2=0$ in which the vortices are embedded.  The monopole zeros are located at the foci of the inscribed hyperbolic 
ellipse.}\label{fig2}
\end{figure}

Recall from section \ref{hyperbolicvortices} that a centered charge $2$ hyperbolic vortex can be constructed from the $C_2$ 
symmetric Blaschke product
\begin{equation}
f(w)\,=\,w\,\frac{w^2+a^2}{1+a^2 w^2},\label{blas}
\end{equation}
where vortex zeros are located at the critical points of $f(w)$ and $a^2$ is related to $\gamma^2$ by $(\gamma^2+1)(a^2+3)=4$.  
Restricting to the action of $f$ on the boundary, it is established in \cite{DGM02} that $f$ is a surjection and that a point 
$w=w_0$ has exactly $3$ preimages $\{w_1,w_2,w_3\}=f^{-1}(\{w_0\})$, defining an ideal triangle.  The edges of this triangle are spectral lines, a 
fact that is readily checked by direct computation in simple cases, or numerically for more generic values of the parameters.  The 
prescribed Blaschke product \eqref{blas} then generates all of the spectral lines (with $\zeta\in\mathbb{R}$) and hence a family 
of ideal triangles corresponding to the gauge freedom in the JNR data.  It was shown in \cite{Sin06} that the envelope of this 
family of triangles is a hyperbolic ellipse (the locus of points for which the sum of the geodesic distances from the foci is 
constant) whose foci are at the critical points of $f$, i.e.~at the vortex zeros.\footnote{On the other hand, joining the triples 
of points $w_i$ by Euclidean triangles would yield a Euclidean ellipse with foci at $w=\pm\text{i}a$, the `non-zero zeros' of 
$f$, \cite{DGM02}.}  Figure \ref{fig2} shows the hyperbolic ellipse for the monopole with $\gamma^2=\tfrac{1}{4}$.

\subsection{Principal axes and spectral radii}\label{principalaxes}
Atiyah and Hitchin \cite{AH88} observed that there are two spectral lines through the centre of a charge $2$ monopole.  This fact 
is used to define the principal axes of the monopole, which in turn define the Euler angles, as natural coordinates on the moduli 
space.  A similar definition is possible in the hyperbolic case.  Spectral lines through the origin of the hyperbolic ball have 
$\eta=\zeta$.  Taking a configuration of the form \eqref{speccharge2} with $\gamma^2\leq\tfrac{1}{3}$, these spectral lines are 
always contained in the plane $X^3=0$, and coalesce along the $X^2$ axis when $\gamma^2=\tfrac{1}{3}$.  The axis $e_1$ is defined 
as the bisector of the angle between these spectral lines.  The second bisector defines the axis $e_2$, which lies in the plane of 
the JNR poles.  The third principal axis, $e_3$, is parallel to the line of separation of the monopole zeros.

The three spectral radii of a Euclidean $2$-monopole are defined as half the separation between the unique two spectral lines 
parallel to each of the three principal axes, \cite{AH88}.  In the hyperbolic setting we will define the spectral radii as the 
minimal geodesic separation between each pair of spectral lines orthogonal to one of the principal axes.  This gives two of the 
spectral radii as the semi-major and semi-minor axes of the hyperbolic ellipse discussed above:
\begin{equation*}
d_{\pm}\,=\,\cosh^{-1}\left(\frac{2}{\sqrt{3\mp2a^2-a^4}}\right).
\end{equation*}
In section \ref{speclinesthroughzeros} we showed that the only pair of spectral lines which meet the equatorial plane at right angles are those 
through the monopole zeros.  This gives the third spectral radius as half the hyperbolic distance between the zeros,
\begin{equation*}
d_3\,=\,\cosh^{-1}\left(\sqrt{\frac{3+2a^2-a^4}{3-2a^2-a^4}}\right).
\end{equation*}
Atiyah \& Hitchin's observation \cite{AH88} that the three spectral radii define a right-angled triangle also holds in the 
hyperbolic case, i.e.~$\cosh(d_-)\cosh(d_3)=\cosh(d_+)$.  From our description, we see that this fact follows immediately from the 
definition of an ellipse.  The area of this triangle is minimal when $a=2^{2/3}-1$.  Curiously, this corresponds precisely to the 
critical radius at which there is a closed geodesic in Hitchin's metric \cite{Hit95}.

\section{Moduli space}\label{secmod}
Low energy scattering of solitons has successfully been modelled by geodesic motion on the moduli space.  The metric on the moduli 
space is given by the $L^2$ norm of perturbations to the fields, subject to the gauge-fixing constraint that gauge orbits are 
orthogonal to such perturbations.

It is well known that the requisite integral diverges for hyperbolic monopoles, although various alternative metrics have been 
proposed.  Examples are Hitchin's metric on the space of spectral curves \cite{Hit95}, and the $L^2$ metric on the space of 
circle-invariant instantons \cite{FS} (both of which have positive scalar curvature).  We will focus on the metric defined via the 
connection on the boundary of $\mathbb{H}^3$, \cite{BA90,MNS03,BCS15}, and compare this metric to the $L^2$ metric on the moduli 
space of the underlying hyperbolic vortices.  In the charge $1$ case these metrics are both proportional to the underlying 
hyperbolic metric.  We thus focus on vortices and monopoles of charge $2$ and fixed centre of mass.

The centered $2$-vortex metric was computed by Strachan \cite{Str92}.  For vortices located at 
$z=\pm\alpha\text{e}^{\text{i}\theta}$ in the Poincar\'e disk, the gauge condition is
\begin{equation}
2\,\partial_i(\delta a_i)+\text{i}\left(\bar{\phi}\delta\phi-\phi\delta\bar{\phi}\right)\,=\,0\label{Strachangauge}
\end{equation}
and the metric takes the form
\begin{IEEEeqnarray}{rcl}
ds^2\,&=&\,\int\left(\frac{1}{2}\,\delta\phi\,\delta\bar{\phi}+\frac{1}{4}\,g^{ij}\,\delta a_i\,\delta a_j\right)\sqrt{g}\,dx\,dr\nonumber\\
&=&\,\frac{2\pi\alpha^2}{(1+\alpha^2)^2}\left(1+\frac{4\,(1+\alpha^4)}{\sqrt{1+14\,\alpha^4+\alpha^8}}\right)\cdot\frac{4\,(d\alpha^2+\alpha^2d\theta^2)}{(1-\alpha^2)^2}.\label{Strachanmetric}
\end{IEEEeqnarray}
Note that the gauge condition \eqref{Strachangauge} does not allow the variations in the fields to be computed by varying the JNR 
function $\psi$ in the gauge defined through (\ref{Mantongaugea}, \ref{Mantongaugeb}).

\newpage
\subsection{Boundary fields}
In order to define a metric on the hyperbolic monopole moduli space, we consider the fields on the boundary of the hyperbolic 
ball.  In this section, we use coordinates $z=x+\text{i}r=x^4+\text{i}x^1$ with metric \eqref{boundarymetric},
\begin{equation}
ds^2\,=\,\frac{1}{r^2}(dx^2+dr^2).\label{bdymetric}
\end{equation}
The boundary fields are obtained by taking the limit $\rho=0$ and $r=x^1$ in \eqref{A1}-\eqref{A4}:
\begin{equation*}
A_4^0\,=\,\frac{\text{i}}{2}\,a_4\tau_1,\qquad\qquad A_1^0\,=\,\frac{\text{i}}{2}\,a_r\tau_1,\label{Abdy}
\end{equation*}
\begin{equation*}
A_\rho^0\,=\,\frac{\text{i}}{2}\left(\frac{\phi_1}{r}\tau_2+\frac{\phi_2+1}{r}\tau_3\right),\qquad\qquad\Phi^0\,=\,-\frac{\text{i}}{2}\,\tau_1.
\end{equation*}
As the Higgs field tends to a constant, the relevant gauge fixing condition is simply the Coulomb gauge 
$\partial_i(\delta a_i)=0$, which holds identically for fields of the form \eqref{Mantongaugeb},
\begin{equation}
a_x\,=\,-\partial_r\log\psi,\qquad\qquad a_r\,=\,\partial_x\log\psi,\label{axar}
\end{equation}
allowing us to obtain the metric by varying $\psi$.
%The curvature of the connection \eqref{Abdy} is then
%\begin{equation*}
%F_{xr}^0\,=\,\partial_xA_1^0-\partial_rA_4^0+[A_1^0,A_4^0]\,=\,\frac{\text{i}}{2}\left(\partial_x^2+\partial_r^2\right)\log\psi\,\tau_1,
%\end{equation*}
%in agreement with the gauge field \eqref{F41} evaluated at $\rho=0$.
%This is traceless, but it may be possible to extract information from $F^2$.  Holonomies on the boundary are probably not meaningful, as the connection is flat so all loops over the boundary $S^2$ are contractible.
The metric is then defined by
\begin{equation}
ds^2\,=\,\int g^{ij}\,\delta a_i\,\delta a_j\,\sqrt{g}\,dx\,dr\,=\,\int\delta^{ij}\,\delta a_i\,\delta a_j\,dx\,dr,\label{metricformula}
\end{equation}
where $g_{ij}$ is the hyperbolic metric on the boundary.  The gauge potentials \eqref{axar} are simply those of a vortex in the 
hemisphere model of $\mathbb{H}^2$.  However, the lack of a Higgs field contribution and the different gauge condition will give a 
metric different from \eqref{Strachanmetric}.

The moduli space metric \eqref{metricformula} is invariant both under 
gauge transformations and conformal rescalings of the boundary metric \eqref{bdymetric}.  The Coulomb gauge condition 
leaves a residual gauge freedom to multiply $\psi$ by the modulus-squared of a holomorphic function, and we use this to remove 
the poles in $\psi$.  The resulting JNR function can equivalently be obtained from the spectral curve polynomial by setting 
$(\zeta,\eta)=(z,-\bar{z}^{-1})$ and multiplying by $\bar{z}^N$.  We denote the resulting function $h$.

%The metric for $N=1$ was determined in \cite{BCS15}.  Restricting to the plane $r=0$ we use the 't Hooft function
%\begin{equation}
%\psi\,=\,1+\frac{t_2^2}{(x-t_1)^2+r^2}\qquad\leftrightarrow\qquad h\,=\,|z-t_1|^2+t_2^2\,,\label{psionevortex}
%\end{equation}
%giving the metric
%\begin{equation*}
%ds^2\,=\,dt^\mu\,dt^\nu\int\frac{\partial a_i}{\partial t_\mu}\frac{\partial a_i}{\partial t_\nu}\,dx\,dr\,=\,\frac{8\pi}{3\,t_2^2}\left(dt_1^2+dt_2^2\right).
%\end{equation*}
%Using \eqref{vortexJNRformula}, we see the physical significance of $t_1$ and $t_2$ as the location of the vortex zero: $(x^4_0,r_0)=(t_1,t_2)$.  From our previous discussions, the monopole Higgs field has a single zero at $(x^4_0,x^1_0,\rho_0)=(t_1,0,t_2)$.

\subsection{Monopole metric: radial component}
We wish to compare \eqref{Strachanmetric} to the radial component of the metric of two hyperbolic monopoles obtained from lifting 
a charge $2$ hyperbolic vortex to $\mathbb{H}^3$.  To compute the metric for two hyperbolic monopoles whose zeros are in the 
plane $x^1=0$, we take the 't Hooft function
%\footnote{Our choice is equivalent to the $D_2$-symmetric family discussed in \cite{BCS15}, with the identifications
%\begin{equation*}
%z_{\text{BCS}}\,=\,\frac{z+\text{i}}{z-\text{i}},\qquad\gamma^2\,=\,\frac{1+a}{3-a},\qquad\lambda^2\,=\,\frac{4(1-a)}{(1+a)(3-a)},
%\end{equation*}
%where the parameter $a$ lies in the range $a\in[-1,1]$ and the monopole zeros are coincident when $a=0$.}
\begin{equation}
\psi\,=\,1+\frac{\lambda^2}{(x^4-\gamma)^2+r^2}+\frac{\lambda^2}{(x^4+\gamma)^2+r^2},\label{charge2JNR}
\end{equation}
where $r^2=\rho^2+(x^1)^2$ and the poles are fixed to lie on the $x^4$ axis.  A geodesic one-parameter family is obtained by imposing dihedral 
symmetry $D_2$, which requires that $2\lambda^2=\gamma^{-2}-\gamma^2$, and this is centered by the definition of \cite{MNS03}.
%The configuration can be rotated by rotating the JNR poles within the equatorial circle $r=0$.  Two rotated configurations are physically equivalent, so the resulting metric is circularly symmetric.
%That this one-parameter-family is indeed a geodesic was established in \cite{BCS15} by imposing $D_2$ symmetry on the spectral curve.  Note that the monopole fields are not rotationally symmetric about the $\rho$ axis.  This would require the vortex Higgs field to be of the form $|\phi|^2=1-r^2f(r^2+(x^4)^2)$, for a suitable function $f$.  This can only be done if the poles of the 't Hooft function enjoy the same rotational symmetry (i.e.~all are at the origin).

\begin{figure}
\centering
\includegraphics[width=0.5\linewidth]{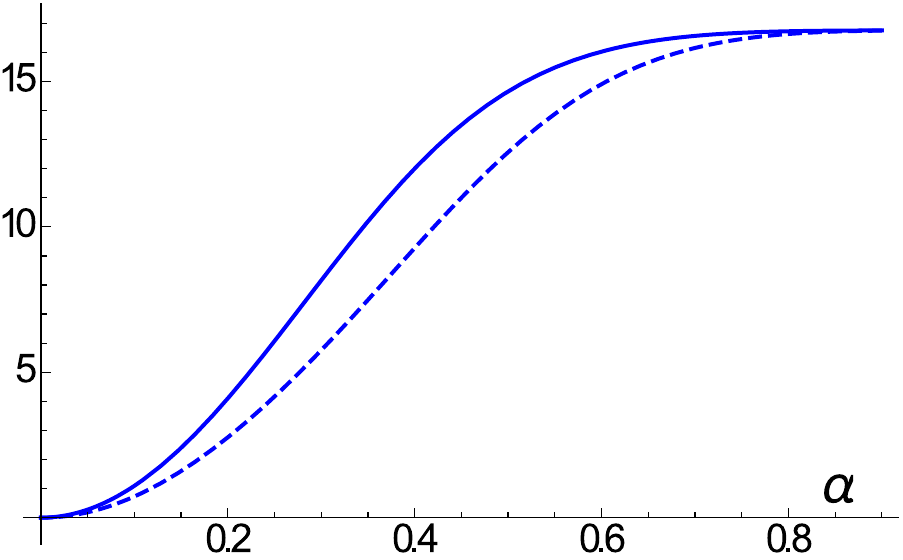}
\caption{Radial component of the metric as a function of $\alpha$, the distance of each Higgs zero from the 
origin.  Solid line:~analytic result \eqref{Strachanmetric} for the vortex metric (rescaled by $32/9$).  Dashed line:~monopole 
metric \eqref{monmet}.  In both cases we have divided by the factor $4(1-\alpha^2)^{-2}$.}\label{fig3}
\end{figure}

To relate $\gamma$ to the positions of the Higgs zeros we must locate, from \eqref{vortexJNRformula}, the zeros of 
$\nabla^2\log(r\psi)$.  There are two regimes:~for $\gamma^2\in[0,\tfrac{1}{3}]$, the zeros are found at $x^4=x^1=0$ and
%\begin{equation}
%\rho^4+2(\gamma^2-\lambda^2)\rho^2+\gamma^2(\gamma^2+2\lambda^2)\,=\,0.\label{JNRzeros}
%\end{equation}
%Requiring $\rho=\rho_0\geq1$ and $\rho=\rho_0^{-1}$ to both solve \eqref{JNRzeros} fixes $\gamma^2(\gamma^2+2\lambda^2)=1$ and
\begin{equation}
\rho_0^{\pm2}\,=\,\frac{1}{2\gamma^2}\left(\left(1-3\gamma^4\right)+\sqrt{\left(1-3\gamma^4\right)^2-4\gamma^4}\right),\label{rhogamma}
\end{equation}
%\begin{equation*}
%\gamma^2\,=\,\frac{1}{6\rho_0^2}\left(\sqrt{(1+\rho_0^4)^2+12\rho_0^4}-(1+\rho_0^4)\right).
%\end{equation*}
while for $\gamma^2\in[\tfrac{1}{3},1]$ they are at $x^1=0$ and
\begin{equation*}
x^4_0\,=\,\pm\,\frac{1}{2\gamma}\sqrt{(1+\gamma^2)(3\gamma^2-1)},\qquad\qquad\rho_0\,=\,\frac{1}{2\gamma}\sqrt{(1-\gamma^2)(3\gamma^2+1)}.
\end{equation*}
Converting back to the ball model of $\mathbb{H}^3$, the monopoles are located at
\begin{equation*}
(X^1,X^2,X^3)\,=\,\left(0,\pm\frac{x^4_0}{\rho_0+1},0\right),\qquad\qquad\frac{1}{3}\leq\gamma^2\leq1
\end{equation*}
or
\begin{equation}
(X^1,X^2,X^3)\,=\,\left(0,0,\pm\frac{\rho_0-1}{\rho_0+1}\right),\qquad\qquad0\leq\gamma^2\leq\frac{1}{3},\label{gammaalpha}
\end{equation}
from which we define
\begin{equation*}
\alpha\,=\,\frac{\rho_0-1}{\rho_0+1}.
\end{equation*}
For ease of numerical computation we recast the JNR function \eqref{charge2JNR} into the form
\begin{equation}
h\,=\,|z|^4-A(\gamma)\,(z^2+\bar{z}^2)+B(\gamma)\,|z|^2+1,\label{hfunction}
\end{equation}
with $A=\gamma^2$ and $B=\gamma^{-2}-\gamma^2$.  We now obtain the radial component of the moduli space metric from 
\eqref{metricformula}, using the relations \eqref{rhogamma} and \eqref{gammaalpha} to change to the coordinate $\alpha$:
\begin{equation}
g_{\alpha\alpha}\,d\alpha^2\,=\,\frac{4\,d\alpha^2}{(1-\alpha^2)^2}\,\rho_0^2\left(\frac{d\gamma^2}{d\rho_0}\right)^2\int\frac{\partial a_i}{\partial(\gamma^2)}\frac{\partial a_i}{\partial(\gamma^2)}\,dx\,dr\,\equiv\,f^2(\alpha)d\alpha^2.\label{monmet}
\end{equation}
%As explained previously, the conformal factor is a surface of revolution.
The integral in \eqref{monmet} is evaluated numerically and the profile function is compared with the (rescaled) metric of the 
corresponding vortex, \eqref{Strachanmetric}, in figure~\ref{fig3}.  Note that in both cases the asymptotic metric approaches that 
of the underlying $\mathbb{H}^3$.
%The full two-dimensional metric of the centered charge $2$ monopole obtained from a charge $2$ vortex (and hence symmetric under $x^1\to-x^1$) is a surface of revolution.  So far we have suppressed the angular coordinate describing a rotation of the three JNR poles, which is reinstated in \eqref{monmet} by the replacements $\alpha^2\to|\alpha|^2$.
%\\ \\
%Now we consider briefly the spectral curve of the two-monopole.  Bolognesi et al \cite{BCS15} give a prescription for determining 
%the spectral curve from the 't Hooft data,
%\begin{equation*}
%\mathcal{S}:\quad\eta^2+\zeta^2+\eta\zeta\left(\gamma^2-\frac{1}{\gamma^2}\right)-\gamma^2\left(\eta^2\zeta^2+1\right)\,=\,0,
%\end{equation*}
%where $\eta$ and $\zeta$ parametrise the space of oriented geodesics in $\mathbb{H}^3$.  Taking a point $\zeta$ on the boundary of 
%the upper half space model of $\mathbb{H}^3$ (with complex coordinate $z=x^4+\text{i}x^1$), the spectral curve picks out a 
%geodesic (a semicircle in the upper half space) to the point $-1/\bar{\eta}$, the antipodal point to $\eta$.  These geodesics pass 
%close to the centre of the monopole.

%Recall that there is a geodesic through each monopole, along which the Higgs field is that of a single hyperbolic monopole.  In 
%the upper half space model, these geodesics are semicircles with centers on the $x^4$-axis and parallel to the $x^1$-axis.  One 
%can easily verify that for any lifted vortex configuration these geodesics are in fact spectral lines with the special property 
%that they go through the monopole zero.

\subsection{Monopole metric: angular components}
\begin{figure}
\begin{minipage}{0.32\linewidth}
\centering
\vspace{-0.3cm}
\includegraphics[width=\linewidth]{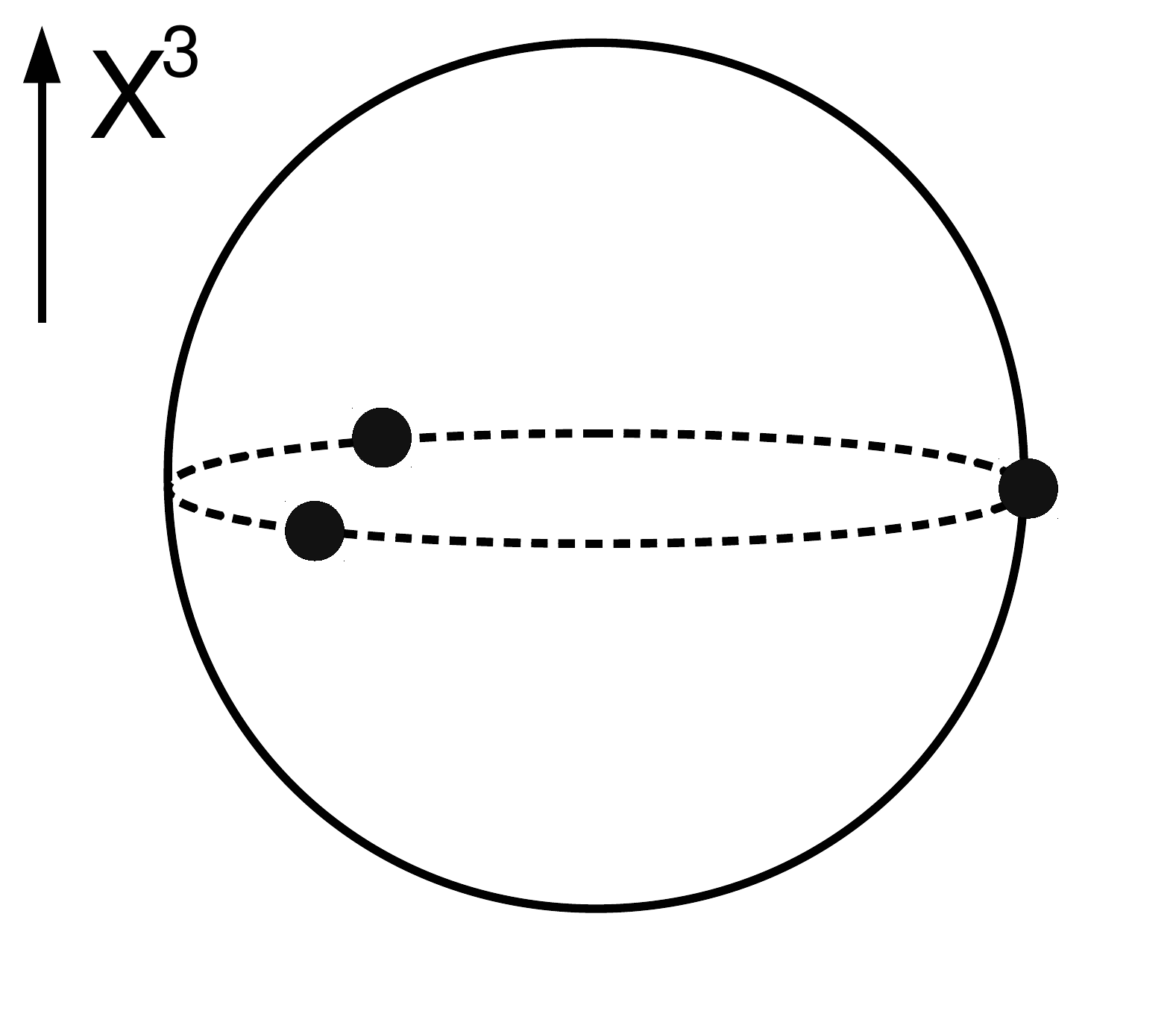}\vspace{-0.3cm}%{frontchain.pdf}
\end{minipage}
\begin{minipage}{0.32\linewidth}
\centering
\vspace{-0.3cm}
\includegraphics[width=\linewidth]{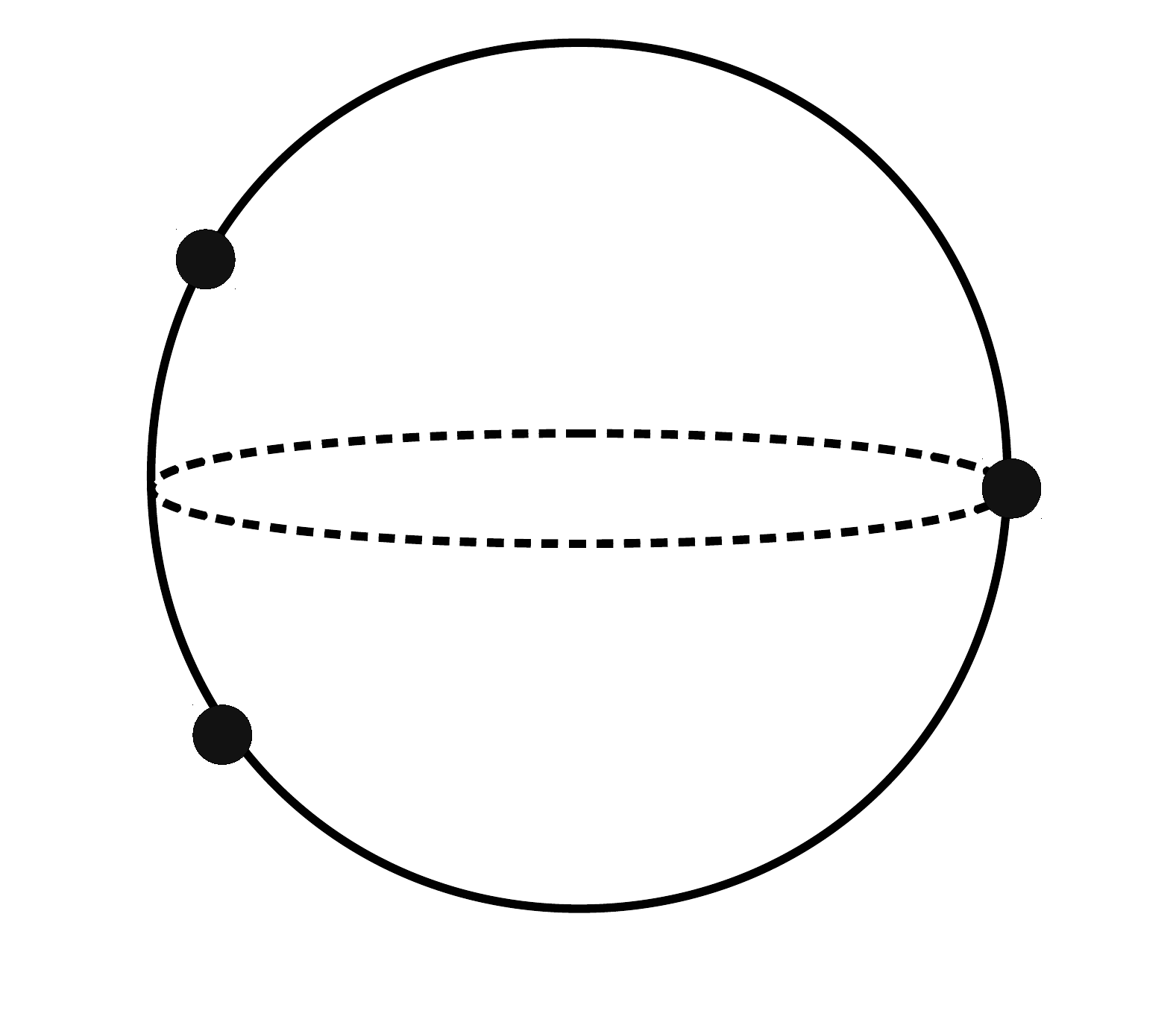}\vspace{-0.3cm}%{frontchain.pdf}
\end{minipage}
\begin{minipage}{0.32\linewidth}
\centering
\includegraphics[width=\linewidth]{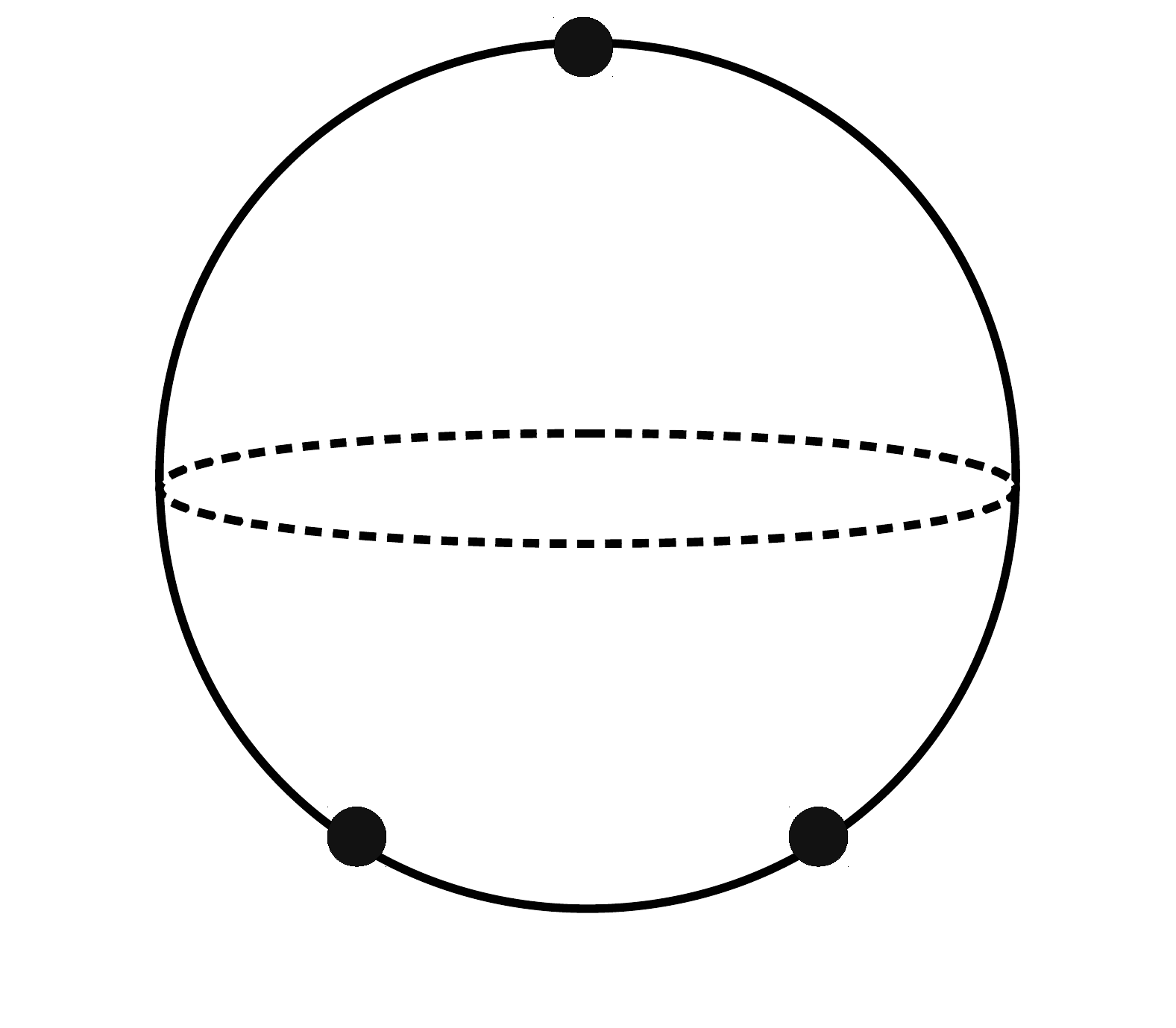}
\end{minipage}\vspace{-0.3cm}
\caption{Orientation of the JNR poles used to compute the functions $a^2$, $b^2$ and $c^2$.  The body-fixed $e_1$, $e_2$ and $e_3$ 
axes are aligned with the spatial $X^3$ axis in turn.}\label{fig4}
\end{figure}

\begin{figure}
\centering
\includegraphics[width=0.5\linewidth]{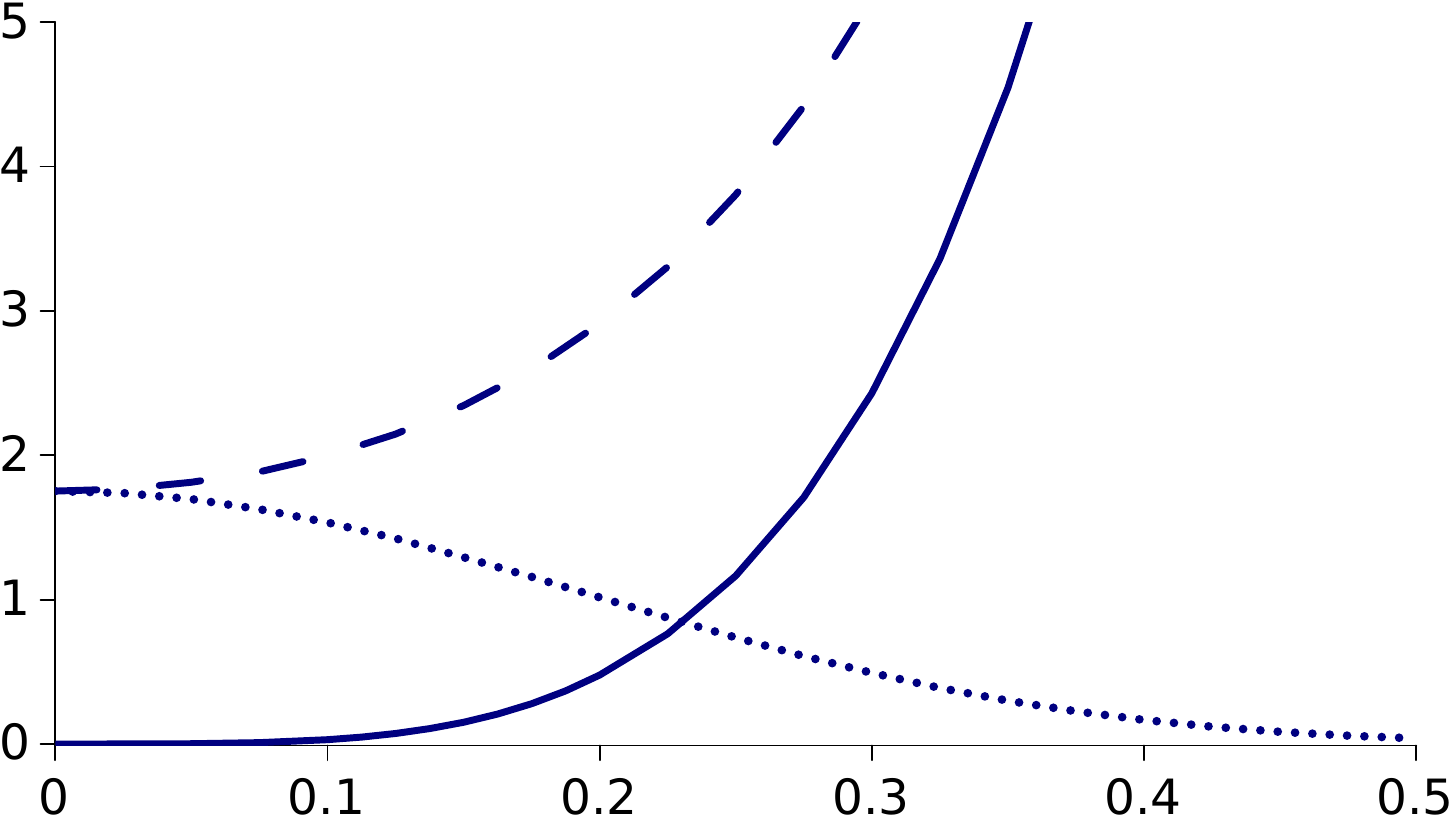}
\caption{Angular component of the metric as a function of $\alpha$.  Solid line:~$a^2(\alpha)$, dashed line:~$b^2(\alpha)$, dotted 
line:~$c^2(\alpha)$.}\label{fig5}
\end{figure}

$\mathrm{SO}(3)$ and dihedral symmetry imply that the moduli space metric of two hyperbolic monopoles is diagonal when expressed 
in terms of the $\mathrm{SO}(3)$-invariant one-forms $\sigma_i$, \cite{AH88}:
\begin{equation*}
g\,=\,f^2(\alpha)\,d\alpha^2+a^2(\alpha)\,\sigma_1^2+b^2(\alpha)\,\sigma_2^2+c^2(\alpha)\,\sigma_3^2.
\end{equation*}
The function $f(\alpha)$ was defined in the previous section.  To compute $a$, $b$, $c$ we rotate the poles of the standard JNR 
function \eqref{charge2JNR} so as to align each of the principal axes $e_1$, $e_2$, $e_3$ (identified in section 
\ref{principalaxes}) with the $X^3$ coordinate axis in turn, as shown in figure \ref{fig4}.
The gauge potential is still determined by functions of the form \eqref{hfunction}, with
\begin{IEEEeqnarray*}{rclcl}
a^2&\qquad\qquad&A\,=\,\frac{1-3\gamma^2}{1+\gamma^2}&\qquad\qquad&B\,=\,\frac{8\gamma^2(1-\gamma^2)}{(1+\gamma^2)^2}\\
b^2&\qquad\qquad&A\,=\,\frac{1-\gamma^2}{1+3\gamma^2}&\qquad\qquad&B\,=\,\frac{8\gamma^2(1+\gamma^2)}{(1-\gamma^2)(1+3\gamma^2)}\\
c^2&\qquad\qquad&A=\gamma^2&\qquad\qquad&B\,=\,\frac{1-\gamma^4}{\gamma^2}.
\end{IEEEeqnarray*}
Deformations are now parametrised by a rotation by an angle $\omega$ in the $z$ plane (which is a stereographic projection of the 
boundary of the unit ball).  This choice of parametrisation fixes the gauge freedom in the JNR data, and $\omega$ represents a rotation about one 
of the principal axes.  For each choice of $A$ and $B$ the relevant component of the metric is given by the integral
\begin{equation*}
\int\left[\left(\partial_\omega a_x\right)^2+\left(\partial_\omega a_r\right)^2\right]\,dx\,dr\,=\,64\,A^2\int\left[\left(\partial_x\left(\frac{rx}{h}\right)\right)^2+\left(\partial_r\left(\frac{rx}{h}\right)\right)^2\right]\,dx\,dr.
\end{equation*}
Plots of these functions are given in figure \ref{fig5}.
Expanding near $\alpha=0$ gives the metric coefficients
\begin{IEEEeqnarray*}{lcl}
f^2\,=\,c_1\,\alpha^2+\mathcal{O}(\alpha^6),&\qquad&b^2\,=\,c_2+c_3\,\alpha^2+\mathcal{O}(\alpha^4),\\
a^2\,=\,\alpha^2f^2+\mathcal{O}(\alpha^6),&\qquad&c^2\,=\,c_2-c_3\,\alpha^2+\mathcal{O}(\alpha^4).
\end{IEEEeqnarray*}
where
\begin{equation*}
c_1\,=\,\frac{32\pi}{9}\,\left(81-10\sqrt{3}\,\pi\right),\qquad c_2\,=\,\frac{8\pi}{27}\,\left(2\sqrt{3}\,\pi-9\right),\qquad c_3\,=\,\frac{4\pi}{9}\,\left(8\sqrt{3}\,\pi-27\right).
\end{equation*}
A numerical computation of the coefficients for $\alpha\to1$ is well fitted by the expansions
\begin{IEEEeqnarray*}{rcl}
f^2\,&=&\,\frac{16\pi}{3}\,\frac{4}{(1-\alpha^2)^2}\left(1-4\,(1-\alpha)^4+\dots\right),\\
a^2\,=\,b^2\,&=&\,\frac{16\pi}{3}\,\frac{4\,\alpha^2}{(1-\alpha^2)^2}\left(1-(1-\alpha)^2+\dots\right),\qquad c^2\,\propto\,(1-\alpha)^8,
\end{IEEEeqnarray*}
which, when $\alpha$ is converted to a hyperbolic distance, is exponentially close to the background hyperbolic metric.  It would 
be interesting to find a physical interpretation of this metric in terms of the forces between well separated monopoles, akin to 
Manton's results in the Euclidean case, \cite{Man85}.  Note in particular that the factor $16\pi/3$ is twice the value obtained 
for a single monopole and plays the role of a mass.  It may also be possible to describe our asymptotic metric by LeBrun's 
hyperbolic analogue \cite{LeB91} of the Gibbons-Hawking metric.

\section{Periodic monopoles}\label{chainssection}
The original motivation for this work was to obtain new examples of hyperbolic monopoles.  The method presented in section 
\ref{symminst} is particularly useful for periodic arrays of monopoles, for which the JNR and ADHM constructions are not currently 
known.  However, periodic and large charge vortex configurations have been studied \cite{MR10,Sut12,MM15} and they are easily 
lifted to $\mathbb{H}^3$.

Periodic monopoles in Euclidean space have previously been studied in some depth via the Nahm transform and spectral curve 
\cite{CK01}.  These tools demonstrated the splitting of the monopole into constituents \cite{War05} and allowed a study of the 
moduli space dynamics \cite{MW13}.

\begin{figure}
\begin{minipage}{0.32\linewidth}
\centering
\vspace{-0.3cm}
\includegraphics[width=0.85\linewidth]{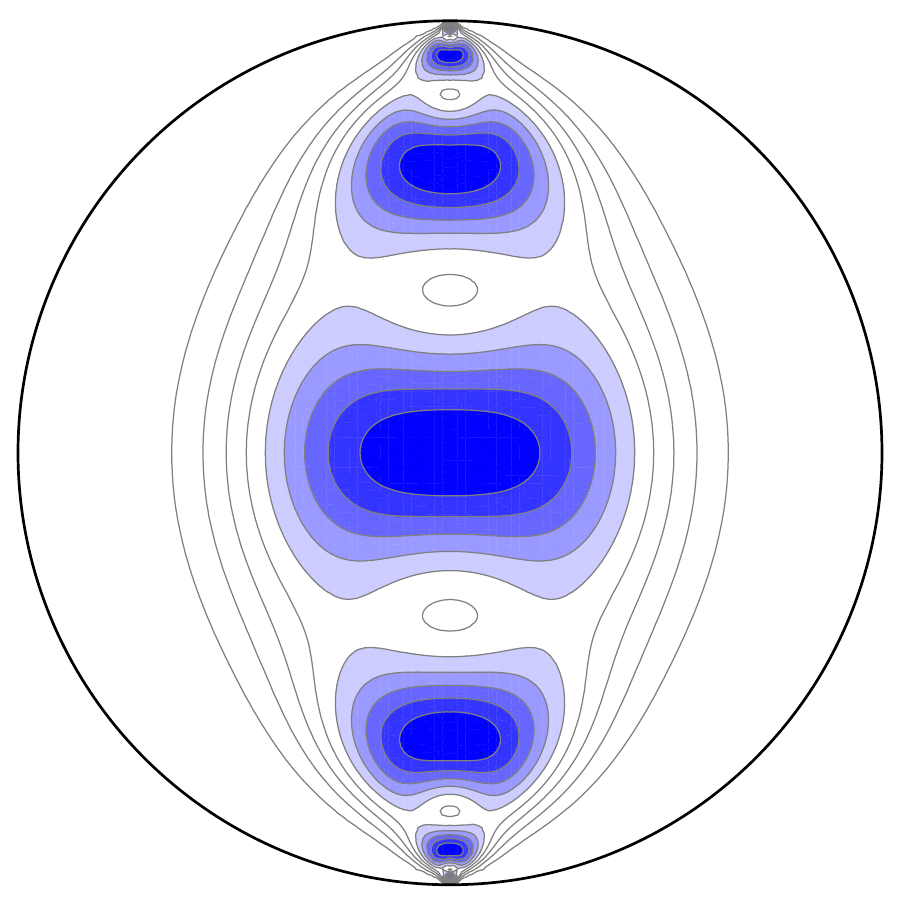}\vspace{-0.3cm}%{frontchain.pdf}
\end{minipage}
\begin{minipage}{0.32\linewidth}
\centering
\vspace{-0.3cm}
\includegraphics[width=0.85\linewidth]{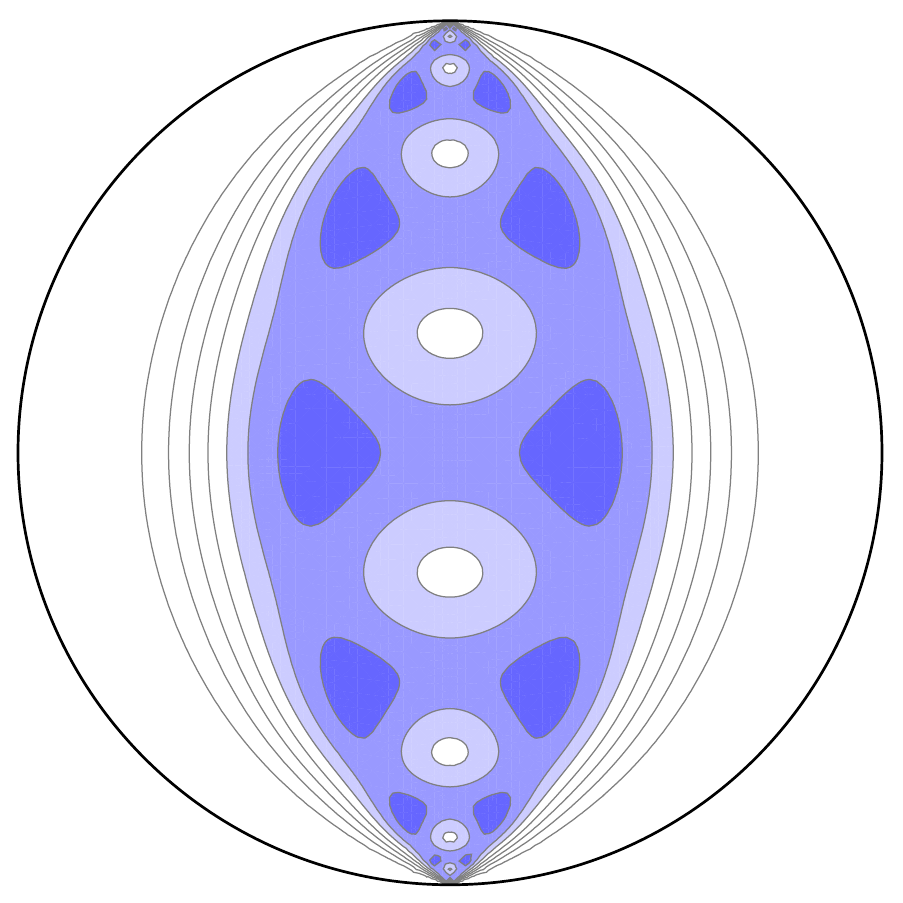}\vspace{-0.3cm}%{frontchain.pdf}
\end{minipage}
\begin{minipage}{0.32\linewidth}
\centering
\includegraphics[width=0.85\linewidth]{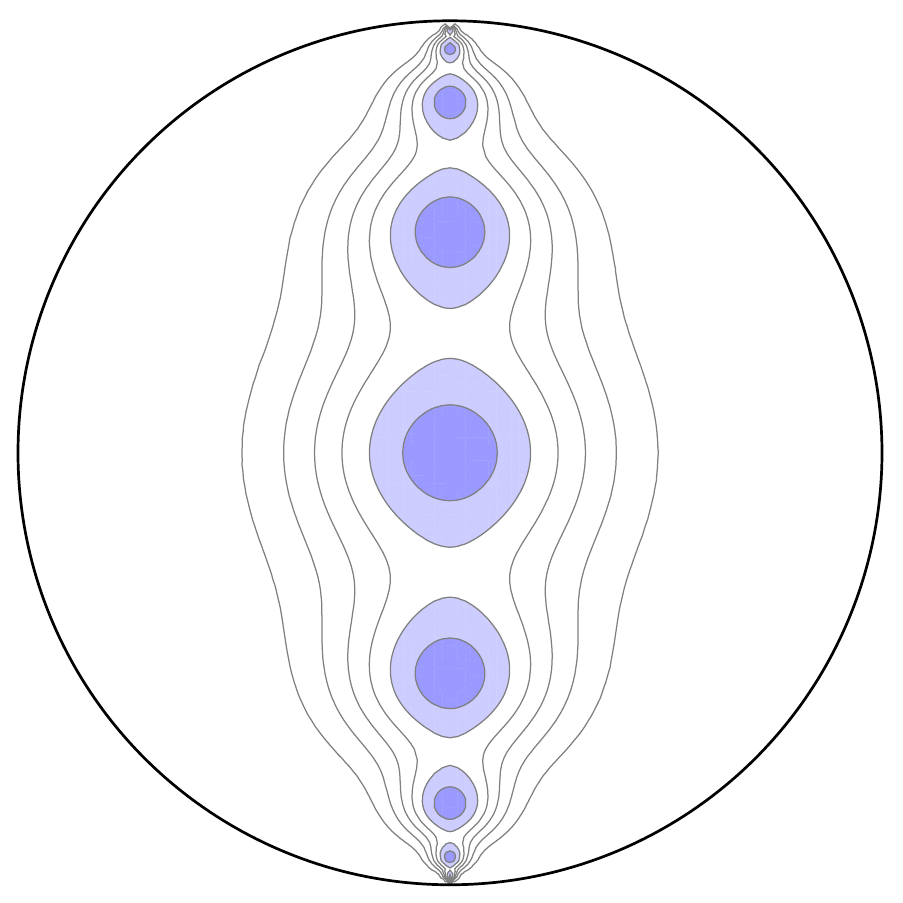}
\end{minipage}
\caption{Slices through a periodic hyperbolic monopole constructed from \eqref{MRchain}.  Contours show energy density in 
intervals of $0.1$, with regions of $\mathcal{E}>0.5$ shaded.  {\bf Left:}~$X^2=0$ plane with $k=0.7$.  The monopoles are 
localised at the zeros of $\Phi$.  {\bf Centre:}~$X^2=0$ plane with $k=0.9$.  Each monopole has split into two constituents, and 
there is a saddle point in energy density at the monopole zeros.  {\bf Right:}~$X^1=0$ plane with $k=0.9$.  Constituents do not 
develop in the $X^2$ direction.}\label{fig6}
\end{figure}

The periodic monopole we will construct in this section is the one lifted from a vortex on the hyperbolic cylinder \cite{MR10}, in 
which the Higgs zeros are strung along a geodesic in $\mathbb{H}^2$.  The JNR data for this periodic vortex is not known, so the 
formula \eqref{modphisquared} provides a novel example of a hyperbolic monopole.  The vortex Higgs field is given in terms of 
elliptic functions, where the elliptic modulus $k$ determines the periodicity.  Explicitly, we use the formula \eqref{phifformula} 
with
\begin{equation}
f(w)\,=\,\frac{\text{cd}_k(2\kappa\tan^{-1}(w))-1}{\text{cd}_k(2\kappa\tan^{-1}(w))+1}\,,\label{Jacobif}
\end{equation}
where $\pi\kappa=2\bK_k$.  Using the coordinate $\text{i}u=\log(x^4+\text{i}r)$ in the upper half space model of $\mathbb{H}^2$ 
gives the Higgs field
\begin{equation}
|\phi|^2\,=\,\kappa^2\left|\text{cn}_k(\kappa u)\text{dn}_k(\kappa u)\right|^2\left(\frac{\text{sin}(\Re(u))}{\Re(\text{sn}_k(\kappa u))}\right)^2.\label{MRchain}
\end{equation}
The monopole constructed from this vortex has zeros at 
$x^4_0=x^1_0=0$ and $\rho_0=\text{e}^{n\lambda/2}$, with $n\in\mathbb{Z}$ and
\begin{equation*}
\lambda\,=\,\frac{\pi\bK'_k}{\bK_k},
\end{equation*}
where $\tfrac{1}{2}\lambda$ is the hyperbolic distance between neighbouring zeros of the Higgs field.  The energy density is 
computed using \eqref{energy} and plotted in figure \ref{fig6} for various values of $k$.
For small $k$ the monopoles are well separated and the energy density is peaked at the Higgs zeros.  As $k$ is increased the 
monopoles get closer together and widen in the $X^1$ direction.  Then at some critical value the energy peaks break apart and 
move away from the $X^3$ axis, leaving the positions of the Higgs zeros as saddle points of energy density.
%The graphs in figure \ref{graphs} show the distance of the energy peaks from the Higgs zeros, as well as the maximum, minimum and saddle point energy densities, as functions of $k$.
%\begin{figure}
%\begin{minipage}{0.485\linewidth}
%\centering
%\vspace{-0.3cm}
%\includegraphics[width=0.9\linewidth]{hypdistk.pdf}\vspace{-0.3cm}%{frontchain.pdf}
%\end{minipage}
%\begin{minipage}{0.485\linewidth}
%\centering
%\includegraphics[width=0.9\linewidth]{endensk.pdf}
%\end{minipage}
%\caption{Graphs illustrating the appearance and behaviour of the constituents of a hyperbolic monopole chain as the period is varied.\\
%{\bf left:}~The solid curve is a graph of $\tfrac{1}{2}\lambda$, the hyperbolic distance between Higgs zeros.  The dashed curve is the 
%hyperbolic distance between an energy maximum and a Higgs zero, computed numerically.\\
%{\bf right:}~Solid curves show the energy density maxima and minima along the chain axis, \eqref{emaxsad}, while the dashed curve is the global 
%energy density maximum.}\label{graphs}
%\end{figure}
\par Expansions of the Higgs field \eqref{MRchain} at the half period points of the periodic vortex (in the Poincar\'e disk model) were 
given in \cite{MM15}.  Applying the formula \eqref{energy} to these expansions yields explicit expressions for the maximal and 
minimal values taken by the Higgs field along the $X^3$ axis:
\begin{equation*}
\mathcal{E}_{\text{min}}\,=\,\frac{1}{2}\left(1-\left(\frac{2\bK_k}{\pi}(1-k)\right)^2\right)^2,\qquad\qquad\mathcal{E}_{\text{saddle}}\,=\,\frac{1}{2}+\left(\frac{2\bK_k}{\pi}\sqrt{1-k^2}\right)^4.\label{emaxsad}
\end{equation*}
The critical value of $k$ at which the maximum energy on the $X^3$ axis becomes a saddle point is found by expanding $|\phi|^2$ to 
higher order.  Performing this expansion and converting back to the upper half-plane model, the energy density at 
$x^4+\text{i}r=\text{i}+|\delta|\text{e}^{\text{i}\theta}$ restricted to $x^1=0$ is
\begin{equation}
\left.\mathcal{E}\right|_{x^1=0}\,=\,\left(\frac{1}{2}+a^2\right)+3\,a|\delta|^2\left(b\cos(2\theta)-a\right)+\mathcal{O}(|\delta|^3),\label{Eexpansion}
\end{equation}
where $a$ and $b$ are the coefficients at order $w^2$ and $w^4$ in an expansion of \eqref{Jacobif}, and are given by
\begin{equation*}
a\,=\,-\frac{4\bK_k^2}{\pi^2}(1-k^2),\qquad\qquad b\,=\,-\frac{8\bK_k^2}{3\pi^2}(1-k^2)\left(\frac{4\bK_k^2}{\pi^2}(1+k^2)-1\right).
\end{equation*}
When $\theta=0$, the order $|\delta|^2$ term in the expansion changes sign when $a=b$, i.e.
\begin{equation*}
8\bK_{k_0}^2(1+k_0^2)\,=\,5\pi^2\qquad\Rightarrow\qquad k_0\,\approx\,0.780.
\end{equation*}
Considering the next term in the expansion \eqref{Eexpansion} shows that the hyperbolic distance of the energy peaks from the 
$X^3$ axis grows like $d\propto(k-k_0)^{1/2}$.  Differentiating \eqref{energy} with respect to $x^1$ shows that the plane 
$x^1=0$ is everywhere a stationary point of the energy density, with a local maximum at the vortex zero.  A similar splitting can 
be observed for chains of finite length.  However, such a splitting has not been observed in the periodic monopole obtained from 
the axially symmetric Harrington-Shepard periodic instanton \cite{HS78}, which gives Higgs zeros equally spaced on a horocycle.

%The splitting into constituents is not only a property of infinite chains but also occurs for chains of finite period, at similar 
%monopole separations.  The infinite monopole chain considered here is generated by the holomorphic function \eqref{Jacobif}, which 
%can be expressed as the infinite product
%\begin{equation}
%f(w)\,=\,w^2\prod_{j=1}^\infty\left(\frac{w^2-a_j^2}{1-a_j^2w^2}\right)^2,\qquad\text{with}\qquad a_j\,=\,\text{i}\,\tanh\left(\frac{j\lambda}{2}\right).\label{series}
%\end{equation}
%In the case considered above, the critical value of $\lambda$ was $\lambda_{k_0}\approx2.842$.  Truncating the series 
%\eqref{series} at $j=1$ gives a chain of five approximately equally spaced monopoles.  Carrying out a similar analysis as before, 
%one finds that constituents appear at $\lambda<\lambda_1$, where
%\begin{equation*}
%\lambda_1\,=\,2\tanh^{-1}\left(\frac{1}{2}\sqrt{\sqrt{17}-1}\right)\,\approx\,2.784.
%\end{equation*}

\section*{Acknowledgements}
Many thanks to Derek Harland, Nick Manton and Paul Sutcliffe for useful comments.  This work was supported by the UK Science and 
Technology Facilities Council, grant number ST/J000434/1.

{\small
\raggedright

}

\end{document}